\documentclass[11pt,a4paper]{article}
\usepackage[utf8]{inputenc}
\usepackage[margin=1in]{geometry}
\usepackage{amsmath,amssymb,amsthm,mathtools}
\usepackage{graphics}
\usepackage{hyperref}
\usepackage{cite}
\usepackage{authblk}
\usepackage{physics}
\usepackage{todonotes}
\usepackage{bbm}
\usepackage{tikz}
\usetikzlibrary{arrows}
\usepackage{xspace}
\usepackage{caption,subcaption}
\usepackage{qcircuit}
\usepackage{comment}
\usepackage{multirow}
\newcommand{\targrel}{*!<0em,.025em>-=-<.2em>{\blacksquare}\qw}

\usetikzlibrary{decorations.pathreplacing}

\newcommand{\ie}{\emph{i.e.}\xspace}

\newcommand{\Id}{\mathrm I}

\newcommand{\ii}{\mathrm i}
\newcommand{\ee}{\mathrm e}

\title{Error mitigation for variational quantum algorithms through mid-circuit measurements}
\author[1,5]{Ludmila Botelho\thanks{lbotelho@iitis.pl}}
\author[1,4]{Adam Glos}
\author[1,5]{Akash Kundu}
\author[1]{Jaros{\l}aw Adam Miszczak}
\author[1,4]{{\"Ozlem Salehi}}
\author[2,3,4]{Zolt\'an Zimbor\'as}
\affil[1]{Institute of Theoretical and Applied Informatics, Polish Academy of Sciences, Ba{\l}tycka~5, 44-100 Gliwice, Poland}
\affil[2]{Wigner Research Centre for Physics, H-1525, P.O.Box 49, Budapest, Hungary}
\affil[3]{BME-MTA Lend\"ulet Quantum Information Theory Research Group, Budapest, Hungary}
\affil[4]{QWorld Association, \url{www.qworld.net}}
\affil[5]{Joint Doctoral School, Silesian University of Technology, Akademicka 2a, 44-100 Gliwice, Poland}
\date{}

\begin{document}
	\maketitle
	
\begin{abstract}
Noisy Intermediate-Scale Quantum (NISQ) algorithms require novel paradigms of error mitigation. To obtain noise-robust quantum computers, each logical qubit is equipped with hundreds or thousands of physical qubits. However, it is not possible to use memory-consuming techniques for current quantum devices having at most hundreds or best thousands of physical qubits on their own.
For specific problems, valid quantum states have a unique structure as in the case of Fock states and W-states where the Hamming weight is fixed, and the evolution takes place in a smaller subspace of the full Hilbert space. With this pre-knowledge, some errors can be detected in the course of the evolution of the circuit, by filtering the states not obeying the pattern through post-selection. In this paper, we present mid-circuit post-selection schemes for frequently used encodings such as one-hot, binary, gray, and domain-wall encoding. For the particular subspace of one-hot states, we propose a method that works by compressing the full Hilbert space to a smaller subspace, allowing projecting to the desired subspace without using any ancilla qubits. We demonstrate the effectiveness of the approach for the Quantum Alternating Operator Ansatz algorithm. Our method is particularly suitable for the currently available hardware, where measuring and resetting are possible, but classical conditional operators are not.
\end{abstract}

\section{Introduction}

The paradigm of Noisy Intermediate-Scale Quantum (NISQ) \cite{preskill2018quantum,bharti2021noisy, kottmann2021tequila} computation is currently of great interest as the current quantum devices are still small, fragile, and prone to noise. Bearing this in mind, NISQ algorithms are designed to use a limited amount of resources to reduce the effect of errors. One strategy is to encompass quantum devices to partially solve the task while utilizing classical computers for the remaining computation. Such hybrid classical-quantum algorithms employ the advantage of having shallow circuits, thus reducing the effect of noise.

One can mention a broad class of Variational Quantum Algorithms (VQAs), in which a cost function is evaluated in the quantum circuit whose parameters are optimized by a classical procedure. Typically for such algorithms, the goal is to find a quantum state $\ket{\psi}$ so that the energy of a
predefined Hamiltonian $\bra\psi H\ket \psi$ is minimized \cite{Cerezo2021}. Variational Quantum Eigensolver (VQE) \cite{peruzzo2014variational} and Quantum Approximate Optimization Algorithms (QAOAs) \cite{farhi2014quantum} are among the most prominent examples of VQAs. The application areas cover an extensive range including chemistry \cite{arute2020hartree}, machine learning \cite{mitarai2018quantum}, circuit compilation \cite{khatri2019quantum}, and classical optimization \cite{moll2018quantum}. 

Although VQA are known to be resilient against coherent errors due to their variational nature \cite{o2016scalable,mcclean2016theory}, the quality of the results is presumably reduced with the impact of decoherence. In the NISQ era, due to a large number of qubit requirement, it is unlikely to utilize quantum error correction methods with VQAs to overcome the effect of noise. Yet, there are various quantum error mitigation (QEM) techniques suitable for the NISQ era \cite{bharti2021noisy,endo2021hybrid}. QEM aims not to recover the ideal quantum state but the ideal measurement outcome through post-processing the measurement results. 

One source of error that limits the current capability of quantum devices is the readout error caused by the imperfect measurement devices. Suppose that we are
interested in the probabilistic distribution  $p:\{0,1\}^n
\to [0,1]$ that results from measuring an $n$-qubit state. Instead of the ideal distribution $p$, the outcome is a stochastically malformed distribution $\mathcal{S}\cdot p$, where $\mathcal{S}$ is a stochastic matrix. Many proposed work \cite{bravyi2021mitigating,geller2020efficient,maciejewski2021modeling} focuses either on the construction of a noise model or to mitigate the noise by classical post-processing through the efficient application of pseudo-inverse $\mathcal{S}^{-1}\mathcal{S}p$. However, the measurement error mitigation is not always sufficient as the errors arising from the evolution of the circuit also play a role. Nevertheless, let us suppose that by the construction of the algorithm, the evolution takes place only on a subspace of full $n$-dimensional Hilbert space and measurement outcomes can be classified as valid or invalid depending on this. As invalid outcomes appear due to the effect of noise, removing them would improve the overall fidelity of the measurement statistics. Classical post-selection performed after the measurement detects some of the evolution errors, but it cannot comprehend the complicated behaviour of the noise acting on the circuit. 

In fact, for all the algorithms mentioned above, a stronger assumption can be proposed: not only the final quantum state is a superposition of the valid states, but the same is true for the quantum state of the circuit during the evolution. Detection and removal of such samples during the circuit implementation can mitigate the errors that arisen through the evolution. Hence, provided that the evolution takes place on a subspace of the whole Hilbert space as in the case of VQE \cite{peruzzo2014variational} and Quantum Alternating Operator Ansatz (QAOA+) \cite{hadfield2019quantum,wang2020x,bartschi2020grover} algorithms, many of the measurement outcomes can be marked as invalid and removed since the error-free computation would never produce them. The possibility of post-selection happens particularly often in quantum physics \cite{arute2020hartree}, chemistry \cite{gard2020efficient,ryabinkin2018constrained} and in principle can be used to classical problems \cite{kerenidis2021classical}. The idea of post-selection performed in the middle of the circuit based on valid states is proposed in \cite{mcardle2019error} for VQE on Hartree-Fock states with fixed Hamming weight. Valid states are filtered by a circuit that computes the electron number, but no numerical implementation results are provided. A similar scheme is proposed in \cite{shaydulin2021error} in the scope of QAOA but only for a restricted type of problems with objective functions preserving symmetries. In \cite{streif2021quantum}, the authors study the effect of depolarizing noise analytically for quantum circuits with particle number conservation. In particular, they focus on QAOA+ with XY-mixers and the Max-$ k $-Colorable-Subgraph problem and investigate how the probability of staying in the feasible space reduces by noise. In addition, they proposed an error-correction scheme for correcting bit-flip errors. 

In addition to the states with fixed Hamming weight $ k $, various valid subspaces appear in the literature as a result of the selected encoding scheme when dealing with VQE or QAOA. When expressing a problem, one often needs to represent an integer using binary variables. One popular approach is using one-hot encoding \cite{lucas2014ising}, which results in one-hot quantum states corresponding to the states with Hamming weight $ k=1 $. Although the already proposed schemes work for one-hot states as well, whether one can further exploit the special property of those states to obtain more efficient error mitigation schemes is unknown. Another approach is using binary encoding to represent integers and it was recently used to obtain qubit-saving formulations for the Travelling Salesman Problem (TSP) \cite{glos2020space}, graph coloring problem \cite{tabi2020quantum}, quadratic Knapsack problem \cite{tamura2021performance} and Max $ k$-Cut problem \cite{fuchs2021efficient}. Finally, an alternative approach is the domain-wall encoding presented in \cite{chancellor2019domain}, and the authors provide special mixers preserving the valid subspace of quantum states for QAOA.

In this paper, we propose schemes for error mitigation in variational quantum circuits through mid-circuit post-selection. The post-selection is performed by injecting a quantum circuit consisting of both gates and measurements. We consider various valid subspaces obtained through different encodings such as one-hot, $k$-hot, binary, and domain-wall encoding that frequently appear in encoding combinatorial optimization problems and in quantum chemistry. In particular, the scheme we propose for one-hot encoding works by compressing the valid subspace to the smaller subspace of quantum states and differentiates from the known methods. We also demonstrate the effectiveness of our approach with an application to QAOA+ for TSP. The proposed error mitigation schemes are suitable, but not limited to NISQ algorithms in principle. Furthermore, they can be currently employed with mid-circuit measurements, recently provided by quantum computers developed by IBM \cite{IBM} and Honeywell \cite{honeywell}.

The rest of the paper is organized as follows. We start with a background on mid-circuit post-selection and error mitigation schemes in Sec. \ref{sec:back}. In Sec. \ref{sec:schemes}, we present various post-selection methods for different valid subspaces. In Sec. \ref{sec:app}, we present the numerical experiments performed for the TSP problem using QAOA+. We conclude by Sec. \ref{sec:conc} with a discussion on future directions.

\section{Background}\label{sec:back}

In this section, we will discuss the effects of noise on quantum circuits and how error can be mitigated through post-selection performed in the middle of the circuit.

\subsection{Error mitigation scheme}

Let $U$ be a quantum circuit with $n$ qubits and initial state $\ket{\psi_0}$. Suppose we're given that the state $\ket{\psi} = U\ket{\psi_0}$ belongs to the subspace spanned by a particular subset $S \subset \{0,1\}^n$ and can be expressed as follows:
\begin{equation}
\ket{\psi} = \sum_{s \in S} \alpha_s \ket{s}. 
\end{equation}
In general, instead of pure quantum state $\ket{\psi}$, we end up with a mixed state $\varrho$ spanned by the whole Hilbert space due to the effect of noise. The ultimate goal of quantum error mitigation is to make $\varrho$ as close as possible to the ideal state $\ket{\psi}$. 

A simple approach to mitigate noise is through classical
\emph{post-selection} applied on the measurement outcomes. Note that if a measurement outcome is not from $S$ then it can be
discarded. In other words, post-selection relies on the
assumption that the mixed state $\frac{\Pi_S \varrho \Pi_S}{\tr(\Pi_S \varrho \Pi_S)}$ defined as the
quantum state $\varrho$ projected through $\Pi_S=\sum_{s\in S} \ketbra{s}$ is a
more faithful representation of $\ket{\psi}$ than $\varrho$.

Let us call the state $\ket{\psi}$ the correct or ideal state, and any other state will be called incorrect. The states spanned by $ S $ are valid, and the states spanned by the remaining are invalid. We distinguish three orthogonal subspaces defined through projections
$P_{1}\coloneqq \ketbra{\psi}, P_{2}\coloneqq \Pi_S  - P_1$ and $P_3\coloneqq\Id
- \Pi_S$. Those projections can be interpreted as follows: $P_1$ is the
projection onto unknown correct state, which would be measured on the noise-free
machine. $P_2$ is the subspace spanned by the incorrect valid states. Those states are
not detectable by post-selection applied after the measurement in the computational basis.
Finally, $P_3$ is the subspace of invalid states, detectable through the
post-selection. The efficiency of post-selection greatly depends on the overlap
of the noisy state $\varrho$ with these subspaces: if the overlap
$\tr(P_2\varrho)$ is high compared to overlap $\tr(P_1\varrho)$ then we should
not expect significant improvement. On the other hand, overlap with
$\tr(P_3\varrho)$ only influences the number of circuit runs to get a fixed
number of valid samples. Projection $P_1+P_2$ defines the valid subspace
and projection $P_2+P_3$ defines the incorrect subspace.

Let us consider depolarizing noise, which turns the ideal state $\ket{\psi}$
into noisy state $\varrho$. A measurement $\{P_1,\Id-P_1\}$ would give back the
ideal state. Nevertheless, it is unreasonable to expect that such measurement can be
implemented in the middle of the circuit in principle, as this would require information about $\ket{\psi}$. On the other hand, 
performing a measurement $\{\Pi_S, \Id-\Pi_S\}$ seems to be much
more plausible since $S$ is known. Although this is still not simple for an arbitrary $S$, it does not require any information other than $S$. 

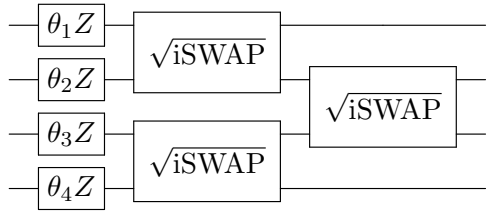
\begin{figure}[h]\centering 
	
	\mbox{\Qcircuit @C=1em @R=1ex{
			& \gate{\theta_1 Z} & \multigate{1}{\sqrt{\rm iSWAP}} & \qw & \qw \\ 
			& \gate{\theta_2 Z} & \ghost{\sqrt{\rm iSWAP}} & \multigate{1}{\sqrt{\rm iSWAP}} & \qw \\
			& \gate{\theta_3 Z} & \multigate{1}{\sqrt{\rm iSWAP}} & \ghost{\sqrt{\rm iSWAP}} & \qw  \\
			& \gate{\theta_4 Z} & \ghost{\sqrt{\rm iSWAP}} & \qw & \qw &\rstick{}
	}}
	\caption{\label{fig:anstatz-subspace-test}Ansatz preserving the subspace of one-hot basis states }
\end{figure}

As an example, suppose that the subspace $ S $ consists of quantum states of Hamming weight 1, so-called one-hot vectors
\begin{equation}
S=\{100\ldots0,010\ldots0,\ldots,000\ldots1\},
\end{equation}
and the valid quantum states are those spanned by $S$.
Let us consider a quantum circuit over $n$ qubits consisting of $l$ layers of the ansatz presented in 
Fig.~\ref{fig:anstatz-subspace-test}. Since the given ansatz does not change the
Hamming weight of the state, starting with a valid state, any obtained quantum state throughout the noiseless evolution of the circuit will belong to the subspace spanned by $S$. The effect of the post-selection discussed above can be improved by performing post-selection in the middle of the circuit by projection onto the subspace $ P_2 $, as the valid states belong to the subspace spanned by $S$ throughout the evolution of the circuit.

\begin{figure}[h]\centering 
	\begin{tikzpicture}[scale=1,->,>=stealth',shorten >=1pt,auto,node distance=3cm,
	thick,main node/.style={font=\sffamily\Large\bfseries}]
	
	\draw[rounded corners] (0, 1) rectangle (3,5) {};
	\draw[rounded corners,dashed] (0.5, 3) rectangle (2.5,4.5) {};
	\draw[rounded corners] (4, 1) rectangle (7,5) {};
	\node (inf) at (5.5,5.5) {\bf invalid};
	\node (feas) at (1.5,5.5) {\bf valid};
	\node (inf) at (1.5,1.8) { incorrect};
	\node (cor) at (1.5,3.75) {correct};
	\node (inf) at (5.5,2.7) {incorrect};	
	\path[every node/.style={font=\sffamily\small}]
	(1.4,3.25) edge [] node [left] {$Z$} (1.4,2.25)
	(1.6,2.3) edge [] (1.6,3.3)
	(2.25,4) edge [] node [above] {$X,Y$} (4.6,4)
	(4.6,3.8) edge [] (2.25,3.8)
	
	(2.6,1.8) edge [] node [above] {$X,Y$} (4.5,1.8)
	(4.5,1.6) edge [] (2.6,1.6)
	;
	
	\end{tikzpicture}
	\caption{\label{fig:post-selection-explanation}A scheme of how $X$, $Y$ and $Z$ errors changes the subspace of the state. }
\end{figure}
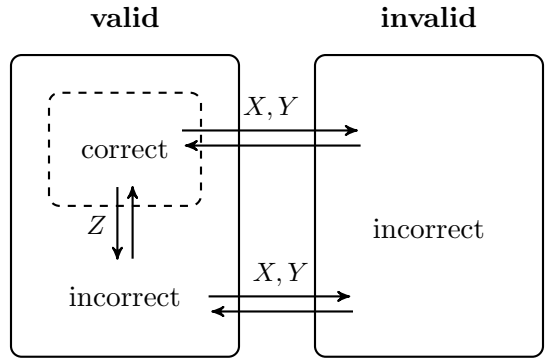

Let us investigate the effect of mid-circuit post-selection in more detail. The evolutions can be roughly decomposed into amplitude transfer between subspaces defined through $P_1,P_2,P_3$, see Fig.~\ref{fig:post-selection-explanation}. If the transition was from valid to invalid states only, then we would not expect any improvement from mid-circuit post-selection compared to the final post-selection. However, the transitions take place also from invalid to valid states. Note that the correct space is only one-dimensional while the dimensionality of the whole valid space usually grows exponentially with the size of the data. Hence, the mid-circuit post-selection attempts to remove the impact of the transitions from invalid states to valid incorrect states mostly.

\subsection{Post-selection by filtering and compression}

Current quantum devices can only measure qubits independently, and thus measurement $\{\Pi_S, \Id-\Pi_S\}$ cannot be applied directly. Even so, we can simulate such measurement. We can distinguish two non-exclusive approaches: \emph{post-selection through filtering} and \emph{post-selection through compression}.

The post-selection through filtering requires ancilla qubits. The idea is to construct a quantum circuit $U_{\rm filter}$ which maps the basis state $\ket{s}$ with $s\in \{0,1\}^N$ such that 
\begin{equation}
U_{\rm filter} \ket{s}\ket{0\cdots 0} = \begin{cases}
\ket{s} \ket{0\cdots 0}, &  s\in S,\\
\ket{s} \ket{\varphi_s}, &  s\not\in S,
\end{cases}
\end{equation}
where $\ket{\varphi_s}$ is (preferably) orthogonal to $\ket{0\cdots 0}$ for any $s\in S$. Upon application of $ U_{\rm filter}  $, the ancilla can be measured in the computational basis and computation continues only if $0\cdots0$ was measured. 

A second approach, post-selection through compression, does not require extra qubits.  Instead, we need a quantum circuit $U_{\rm compress}$, which compresses valid states to a some smaller subspace $S'\subseteq \mathcal H$ such that
\begin{equation}
U_{\rm compress} \ket{s}= \begin{cases}
\ket{\psi_s}\ket{0\cdots 0}, & s \in S,\\
\ket{\varphi_s}, & s \not\in S.
\end{cases}
\end{equation}
Note that here the only requirement is that some qubits are `reset' to $\ket{0}$ after $U_{\rm compress}$. Like in post-selection through filtering, the qubits are measured, and the computation continues \textit{iff} all qubits are in state $ \ket{0} $. In this case, we uncompute the compression through $U_{\rm compress}^\dagger$. An evident advantage of this method compared to the previously introduced one is that it can run in-place without extra qubits. 

The proposed methods are particularly suitable for quantum devices that allow mid-circuit measurements and can reset qubits to $\ket{0}$. Indeed, in this case, the number of required qubits does not grow with the number of applications of the proposed techniques. Still, it is also possible to harness quantum devices without the mid-circuit measurements feature. It is enough to implement filtering each time with different ancilla and to uncompute the state to a new set of qubits in the compression case. Then, the number of additional qubits will be proportional to the number of corrections applied and the number of measured qubits. However, a large number of mid-measured qubits or the number of post-selections applied makes the approach significantly less NISQ-friendly.

It is not possible to provide a general description of how to implement $U_{\rm compress}$ or $U_{\rm filter}$. The reason behind this is that the structure of $S$ depends on the form of the Hamiltonian and the origins of the optimization problem. In the following section, we discuss the implementations of post-selection circuits for different $ S $, which are specifically useful for various combinatorial and physical optimization problems.

\section{Postselection schemes for different encodings}\label{sec:schemes}

For the methods to be NISQ-friendly, they should use as few resources as possible. The resources usually considered are the number of ancilla qubits, the number of gates, and the depth of the circuit. These three, together with the volume, will be our main resources considered in the paper.

Before moving on to the description of specific error mitigation schemes for different encodings, we would like to recall the circuit counting the electron number from  \cite{mcardle2019error}. Since Jordan-Wigner transformation is used where qubits represent spin-orbitals, and occupation number is represented by 0 or 1, counting the electron number is simply counting the number of 1's in a basis state. The circuit described in \cite{mcardle2019error} computes the number of 1's in binary, one bit at a step, using only a single ancilla. The idea can be used as a subroutine in other circuits to verify whether the total number of 1's is a particular value.

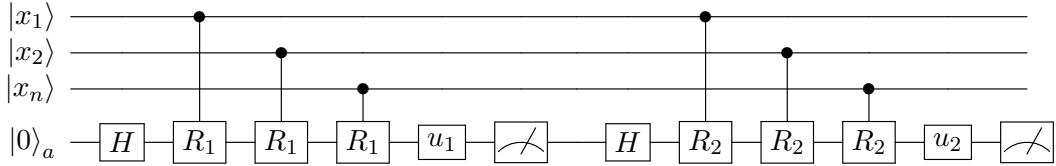
\begin{figure}
	\centering
	\mbox{\Qcircuit @C=1em @R=1em {
			\lstick{\ket{x_1}}	&\qw		&\ctrl{3}	&\qw 		&\qw 		&\qw			&\qw		&\qw		&\qw		&\ctrl{3}	&\qw 		&\qw 		&\qw			&\qw\\
			\lstick{\ket{x_2}}	&\qw		&\qw 		&\ctrl{2}	&\qw 		&\qw			&\qw		&\qw		&\qw		&\qw 		&\ctrl{2}	&\qw 		&\qw			&\qw	\\
			\lstick{\ket{x_n}}	&\qw		&\qw 		&\qw 		&\ctrl{1}	&\qw			&\qw		&\qw		&\qw		&\qw 		&\qw 		&\ctrl{1}	&\qw			&\qw	\\
			\lstick{\ket{0}_a}	&\gate{H}	&\gate{R_1} &\gate{R_1}	&\gate{R_1}	&\gate{u_1}&\meter		&\qw		&\gate{H}	&\gate{R_2} &\gate{R_2}	&\gate{R_2}	&\gate{u_2}&\meter	\\
	}}
	\caption{\label{fig:sum-rotation}An example implementation of the circuit verifying whether the total number of 1's is equal to $\kappa$. The gate $R_j $ is given by diag($ 1,e^{\pi i /2^{j-1}} $) and $ u_j $ is given by diag$(1,e^{-\rm{dec}(\kappa_{j-1}\cdots \kappa_1)\pi i /2^{j-1}})$ where $ \rm{dec}(\kappa_{j-1}\cdots \kappa_1)$ is the decimal representation of the least significant $ j $ bits of the binary string $ \kappa = \kappa_{n}\kappa_{n-1}\cdots \kappa_1 $. $ u_1 $ is defined as the identity operator.}
\end{figure}

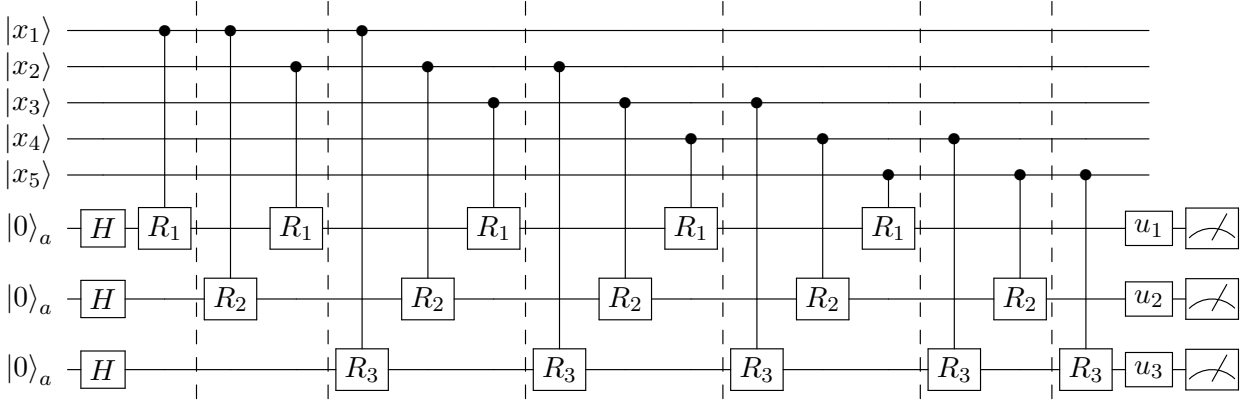
\begin{figure}
	\centering
	{\mbox\small
		\Qcircuit @C=5pt @R=1em {
			\lstick{\ket{x_1}} &\qw &\ctrl{5} \barrier[-2.7ex]{7}&\ctrl{6}& \qw \barrier[-2.7ex]{7} &\ctrl{7} &\qw &\qw \barrier[-2.7ex]{7} &\qw &\qw &\qw \barrier[-2.7ex]{7} &\qw &\qw &\qw \barrier[-2.7ex]{7} &\qw & \qw \barrier[-2.7ex]{7} & \qw &\qw\\
			\lstick{\ket{x_2}} &\qw &\qw & \qw&\ctrl{4} &\qw & \ctrl{5} &\qw &\ctrl{6} & \qw &\qw &\qw &\qw &\qw &\qw&\qw&\qw&\qw\\
			\lstick{\ket{x_3}} &\qw &\qw &\qw &\qw &\qw & \qw &\ctrl{3} &\qw &\ctrl{4} &\qw &\ctrl{5} &\qw &\qw &\qw&\qw&\qw&\qw\\
			\lstick{\ket{x_4}} &\qw &\qw &\qw &\qw &\qw &\qw &\qw &\qw &\qw &\ctrl{2} &\qw &\ctrl{3} &\qw &\ctrl{4} &\qw &\qw &\qw\\
			\lstick{\ket{x_5}} &\qw &\qw &\qw &\qw &\qw &\qw &\qw &\qw &\qw &\qw &\qw &\qw &\ctrl{1} & \qw & \ctrl{2} & \ctrl{3} &\qw \\
			\lstick{\ket{0}_a} &\gate{H} &\gate{R_1} &\qw &\gate{R_1} &\qw&\qw &\gate{R_1}&\qw&\qw &\gate{R_1}&\qw&\qw &\gate{R_1} & \qw & \qw & \qw & \gate{u_1} & \meter \\
			\lstick{\ket{0}_a} &\gate{H} &\qw & \gate{R_2} & \qw &\qw &\gate{R_2} & \qw &\qw &\gate{R_2}& \qw &\qw &\gate{R_2}& \qw &\qw &\gate{R_2} & \qw & \gate{u_2} & \meter \\
			\lstick{\ket{0}_a} &\gate{H} &\qw &\qw &\qw &\gate{R_3} & \qw &\qw &\gate{R_3}& \qw &\qw &\gate{R_3} & \qw &\qw &\gate{R_3} & \qw &\gate{R_3} & \gate{u_3} &\meter \\
	}}
\caption{\label{fig:sum-rotation-parallel}An example implementation of alternative circuit verifying whether the total number of 1's is equal to $\kappa$. The idea behind is the same as the one presented in Fig.~\ref{fig:sum-rotation} except we have $\sim \log n $ ancilla and we apply gates in parallel.}
\end{figure}

Let us describe the verification circuit inspired by \cite{mcardle2019error}. Suppose that we want to verify whether the basis state $ \ket{x_1x_2 \cdots x_n} $ contains exactly $ k$ 1's. Let $ \kappa $ be the binary representation of $ k$ written using $ \lceil \log n \rceil $ bits (0's are padded to the most significant bits if  $ \lceil \log k \rceil < \lceil \log n \rceil $) and let $ \xi $ denote the binary representation of the sum of 1's in $ \ket{x} $. The circuit computes $ \xi $ starting from the least significant bit, as long as the measured bits coincide with that of $ \kappa $'s. In general, for an $ n $-qubit circuit, there are $ \lceil \log n \rceil $ blocks each computing a bit of $ \xi $. After each block, the ancilla qubit is measured in the $ X $-basis. If the measurement result is $ \ket{+}$, then it indicates that the bit is 1 and the measurement result $ \ket{-} $ indicates that the bit is 0. Note that there are two possible outcomes when running the verification circuit: If at some stage the measurement outcome does not coincide with $ \kappa $ the computation ends, or all $ n $ bits coincide indicating that the verification succeeds. We would like to remark that all $ \lceil \log n \rceil $ bits of $ \xi $ should be computed  since it can be the case that $ \kappa $ and $ \xi $ coincide on the first $ \lceil \log k \rceil $ bits, although $ \kappa $ and $ \xi $ are different.

In Fig. \ref{fig:sum-rotation}, a circuit with $ n=3 $ control qubits and a single ancilla qubit is given. Note that there are 2 blocks in the given circuit as the sum can be at most $ 11_2 $. If the first measured bit is not the least significant bit of $ \kappa $, then the computation ends. Otherwise, the computation continues with the second block.

The overall number of required gates and the depth are $ O(n \log n) $. However, one can apply the controlled rotations in parallel, given extra ancilla qubits. The idea for $n=5$ is presented in Fig.~\ref{fig:sum-rotation-parallel}. This approach requires $\sim\log n$ ancilla and the depth equals $O(n)$. Note that in this case each bit of $\xi$ is stored on a different ancilla qubit. 

\subsection{$k$-hot encoding}\label{sec:khot}

The $k$-hot states are 0-1 states with Hamming weight $k$ and often appear in physics and computer science: $k$-hot vectors for $k\geq2$ are a natural description of quantum $k$-particle Fock spaces \cite{arute2020hartree, georgescu2014quantum, mcardle2019error}. Dicke states which are the equal superposition of $ k $-hot states are used as the initial state in QAOA \cite{bartschi2019deterministic} for certain problems. $ k $-hot states are also used to encode the feasible states in problems like Max-$ k$ Vertex Cover problem \cite{cook2020quantum}, and graph partitioning \cite{hadfield2019quantum}.

Post-selection can be applied to $k$-hot states through filtering by verifying the total number of 1's in the quantum state using the circuits given in Fig. \ref{fig:sum-rotation} or \ref{fig:sum-rotation-parallel}. The idea was first investigated in \cite{mcardle2019error}, in the scope of VQE and particle number preserving ansatz.

\subsection{One-hot encoding}\label{sec:compression}

One-hot encoding is a special case of $k$-hot encoding, and it is used in literature for encoding various problems like Travelling Sal esman Problem, Graph Coloring, and Clique Cover \cite{lucas2014ising}. It is also used for optimization over functions $\sigma:\{1,\dots,n\}\to \{1,\dots,m\}$. In the latter case, we specify $ n $ quantum registers, and each register consisting of $ m $ qubits that encode the values of the function between 1 and $ m $.

\begin{figure}\centering
	
	\mbox{\Qcircuit @C=1.5ex @R=2ex {
			\lstick{ } & \qw & \qw & \qw & \qw & \qw & \qw & \qswap & \qw & \qw & \qswap & \targ & \targ & \qw & \rstick{\hspace{-1em}\ket{0}}\\
			\lstick{  } & \qw & \qw & \qw & \qw & \qw & \qswap & \qw & \targ & \qw & \qw & \ctrl{-1} & \qw & \qw & \rstick{\hspace{-1em}\ket{b_{1}}}\\
			\lstick{ } & \qw & \qw & \qw & \qw & \qswap & \qw & \qw & \qw & \ctrl{1} & \qswap \qwx[-2] & \qw & \qw & \qw & \rstick{\hspace{-1em}\ket{0}}\\
			\lstick{ } & \targ & \qw & \qw & \qw & \qw & \qw & \qw & \ctrl{-2} & \targ & \ctrl{-3} & \qw & \qw & \qw & \rstick{\hspace{-1em}\ket{b_{2}}}\\
			\lstick{ } & \qw & \qw & \qw & \ctrl{3} & \qw & \qw & \qswap \qwx[-4] & \qw & \qw & \qw & \qw & \qw & \qw & \rstick{\hspace{-1em}\ket{0}}\\
			\lstick { } & \qw & \qw & \ctrl{2} & \qw & \qw & \qswap \qwx[-4] & \qw & \qw & \qw & \qw & \qw & \qw & \qw & \rstick{\hspace{-1em}\ket{0}}\\
			\lstick{   } & \qw & \ctrl{1} & \qw & \qw & \qswap \qwx[-4] & \qw & \qw & \qw & \qw & \qw & \qw & \qw & \qw & \rstick{\hspace{-1em}\ket{0}}\\
			\lstick{  } & \ctrl{-4} & \targ & \targ & \targ & \ctrl{-5} & \ctrl{-6} & \ctrl{-7} & \qw & \qw & \qw & \qw & \qw & \qw & \rstick{\hspace{-1em}\ket{b_{3}}}\\
	}}
	\caption{\label{fig:enc-change} An example of the implementation of the map $V$ which transforms the one-hot encoding to binary encoding \cite{sawaya2020resource}. Note that the procedure can be adjusted to the case where the maximal stored number is not a power of 2. If an initial state is a superposition of one-hot basis states, then some of the output qubits are set to $\ket{0}$.}
\end{figure}
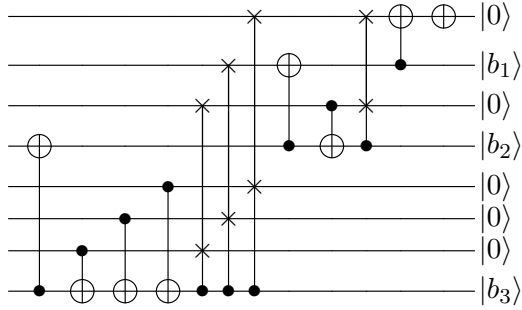

We will mention two different approaches for post-selecting one-hot states. Since one-hot vectors are a special case of $k$-hot vectors with $k=1$, we can use the filtering approach proposed in the Sec. \ref{sec:khot}. Alternatively, one can consider a post-selection through compression with a circuit that converts one-hot representation to binary representation \cite{sawaya2020resource}. Let $V$ be the unitary operation implementing this map. For an integer $ l\in \{1,\dots,n\} $, let $OH_n(l)$ be the bit assignment for one-hot encoding, i.e. it maps $ l $ to the quantum state with a 1 in the $ l $'th position. Let $B_m(l)$ be the bit assignment function encoding $ l $ in binary using exactly $ m $ bits. $B_m$ maps $ l $ to $b_m\dots b_1$ such that $ l = \sum_{i=1}^m 2^{i-1} b_i$. Although the map $V: OH_n(l) \mapsto B_{m}(l)$ does not preserve the number of qubits, the unoccupied qubits after the transformation are set to $\ket{0}$ as it can be seen in Fig.~\ref{fig:enc-change}. The one-hot to binary conversion leaving some qubits in-state $ \ket{0} $ provides a natural scheme for error mitigation. 

The circuit implementing $ V $ uses $ O(n) $ gates, no ancilla, and has $ O(n) $ depth \cite{sawaya2020resource}. After applying $ V $ and measuring the qubits which should be in state $ \ket{0} $, $ V^{\dagger} $ should be applied for decompression. Note that the compression approach uses fewer resources when compared to the $ k$-hot filtering approach for one-hot states.

\subsection{Domain-wall encoding}

In domain-wall encoding~\cite{chancellor2019domain}, valid states are of the form $\ket{1\cdots 10\cdots0}$, \ie the state starts with some number of ones followed by zeros. It requires less connectivity for checking the feasibility condition. For instance, in one-hot encoding, it is required to check whether each pair of qubits are in state $ \ket{1} $ or not, while for domain-wall it is sufficient to check only neighboring qubits to see whether a $ \ket{0} $ is followed by $ \ket{1} $. Note that any problem expressed using $ n $ qubits in one-hot encoding can be also expressed by domain-wall encoding, such that integer $ l $ is represented with a quantum state where $ l$ ones are followed by $ n-l $ zeros.

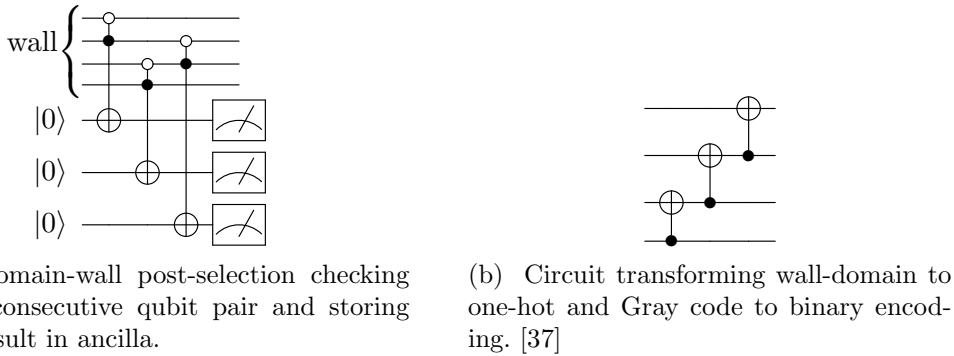
\begin{figure}\centering 
	\begin{subfigure}[b]{0.4\textwidth}\centering
		\mbox{\Qcircuit @C=0.5em @R=.4em {
				& \ctrlo{1} & \qw & \qw & \qw \inputgroupv{1}{4}{.8em}{.8em}{\rm wall} \\
				& \ctrl{3} & \qw &\ctrlo{1} & \qw \\
				&\qw& \ctrlo{1} & \ctrl{4} & \qw \\
				&\qw & \ctrl{2} & \qw & \qw \\
				\lstick{\ket0}	& \targ & \qw &\qw & \meter \\
				\lstick{\ket0}	& \qw & \targ &\qw & \meter \\
				\lstick{\ket0}	& \qw & \qw &\targ & \meter \\
		}}
		\subcaption{\label{fig:domain-wall-post}Domain-wall post-selection checking each consecutive qubit pair and storing the result in ancilla.}
	\end{subfigure}
	\hspace{0.2in}
	\begin{subfigure}[b]{0.4\textwidth}\centering
		\mbox{\Qcircuit @C=0.5em @R=.8em {
				& \qw & \qw & \targ & \qw  \\
				& \qw & \targ & \ctrl{-1} & \qw \\
				& \targ & \ctrl{-1} & \qw & \qw \\
				&\ctrl{-1} & \qw & \qw & \qw \\
		}}
		\subcaption{\label{fig:wall-to-onehot} Circuit transforming wall-domain to one-hot and Gray code to binary encoding.\cite{sawaya2020resource}}
	\end{subfigure}
\caption{\label{fig:wall-domain} Circuits used for postselection for wall-domain encoding.}
\end{figure}

The conditions above also motivate a mid-circuit post-selection scheme through filtering as invalid states can be detected by checking consecutive bits. One approach is to check each neighboring pair of qubits and store the result using $ n-1 $ ancilla. While very demanding in the number of qubits, the approach requires only $ O(1)$ depth and $ O(n) $ gates. An example circuit with 4 qubits can be found in Fig.~\ref{fig:domain-wall-post}. When the number of qubits is limited, then one can apply the error checking with the output on a single ancilla qubit, and measure it instantly and reset it so that it will be reused for the next condition checking. While the number of ancilla qubits will be only one, the depth will increase to $O(n)$.

Finally, one can use an ancilla-free method by first transforming domain-wall to one-hot encoding using the circuit given in Fig.~\ref{fig:wall-to-onehot} and use the post-selection through compression method described in the previous subsection. In this case, the number of gates and depth is the same as for one-hot vectors which are $O(n)$ for both.

\subsection{Binary and Gray encoding}\label{sec:binarytogrey}

\begin{figure}[t]\centering
	\mbox{\Qcircuit @C=0.5em @R=0.5em {
			& \qw & \qw & \qw & \qw & \qw & \qw \\
			& \qw & \qw & \qw &\qw &\ctrl{1}  & \qw  \\
			& \qw & \qw &\ctrl{1} & \qw &\ctrlo{1} & \qw \\
			& \qw & \qw &\ctrl{1} & \qw &\ctrl{1} & \qw \\
			& \ctrl{1} & \qw &\ctrlo{1} & \qw &\ctrlo{1} & \qw \\
			& \ctrl{1} & \qw &\ctrl{1} & \qw &\ctrl{1} & \qw \\
			\lstick{\ket{0}} & \targ & \meter & \targ & \meter &\targ & \meter & \\
	}}
	\caption{\label{fig:binary-filter} Binary exact post-selection for $\mu=42=101001_2$}
\end{figure}
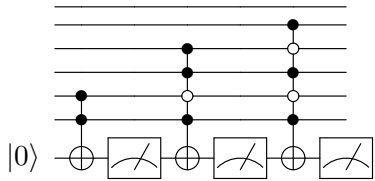

Binary encoding of an integer $ l $ using $ n $ bits is obtained by the assignment $ B_n(l) $ as discussed in Sec. 3.2. It is used in QUBO formulations to save qubits while representing slack variables as discussed in \cite{lucas2014ising}. It is also used while formulating qubit efficient higher-order unconstrained binary optimization formulations (HOBO) for problems like TSP \cite{glos2020space} and graph coloring \cite{tabi2020quantum}.

Using $ n $ bits, the numbers $0,1,\dots,2^{n-1}-1$ are naturally encoded. If not all of the integer values encoded using $ n $ bits are admissible, then some of the encoded integers will be invalid and this will increase the infeasible space. There are several workarounds to solve this issue. One approach is to use the knowledge about the maximal attainable value $\bar x$ \cite{karimi2019practical}, and update the encoding as
\begin{equation}
\sum_{i=1}^{n} 2^{i-1} b_i + \left (\bar x - \sum_{i=1}^{\lceil \log(\bar x )\rceil}2^{i-1} \right ) b_n,
\end{equation}
which introduces bias for higher values. However, when the numbers encoded are the slack variables turning inequality $ f(b) \geq 0 $ into $ f(b) +x_i=0 $, usually small values of $ x_i $ are encountered so that the original inequality is satisfied tightly or almost tightly. For this reason, introducing bias for higher values may have a negative effect on the optimization. Furthermore, in recent HOBO formulations using binary encoding \cite{glos2020space,tabi2020quantum}, quantum states which encode too large values have to be penalized unlike the method above. However one may expect a variation of this algorithm with QAOA+ which will forbid (up to noise) quantum state from evolving into too large numbers, for example, a particular version of QAOA+. Motivated by this, and also for completeness, we describe a filtering scheme below.

Suppose the valid integer can be at most $\mu=b_n'b_{n-1}'\cdots b_1'$. Let $I_0'$ be the collection of indices $i$ for which $b_i'=0$. The bit assignment is invalid ($\ie$ encodes a larger integer than $ \mu $) if at some bit at which it should be zero, it is one \emph{and} all of the more significant bits are the same as that of $\mu$. For example, if we have $\mu=42=101001_2$, then incorrect numbers are of the form $11b_3b_2b_1b_0$, $1011b_1b_0$ and $10101b_0$, where $b_i$ are arbitrary. Hence, we need to verify whether any such situation occurs. An exemplary post-selecting circuit proposed in Fig.~\ref{fig:binary-filter} for $ \mu=42 $. 

In the worst case, for instance when $ \mu = 10\cdots 0_2 $, one may need to check $ n-1 $ invalid forms. Each check requires implementation of multi-controlled NOT gates. To implement a multi-controlled NOT gate controlled by $ n $ qubits, we will consider two different methods: The first method described in \cite{saeedi2013linear} uses no ancilla, requires $O(n^2)$ gates and has $ O(n) $ depth. The second method proposed in \cite{he2017decompositions} uses $ O(n) $ ancilla, $ O(n) $ gates and has $ O(\log n) $ depth.

Hence, if one wants to save ancilla qubits, then the ancilla-free implementation of multi-controlled NOT gate is more convenient, the overall approach requiring a single ancilla qubit, $ O(n^3) $ gates and the circuit has $ O(n^2) $ depth. Using the second method, overall circuit requires $ O(n) $ ancilla, $ O(n^2) $ gates and has $ O(n \log n) $ depth.

In general, checking all invalid forms might be costly depending on the value of $ \mu $ as the error mitigation itself might introduce some errors. For instance, when $ \mu= 10 \cdots 0 $, checking only the most significant bit that should be 0 is enough to eliminate half of the invalid cases. In general, this would require only a single application of multi-controlled NOT gate, and in the worst case when $ \mu = 11\cdots 10 $ there will be $ n-1 $ control qubits. In such a case, it may not be efficient to use an error mitigation circuit only to eliminate a single invalid state. However, if there are multiple registers, say $k$,  encoding numbers in binary, then the proportion of the feasible to all states equal
\begin{equation}
\left(\frac{n-1}{n}\right)^k \approx \ee ^{-\frac{k}{n}}.
\end{equation}
Even for this extreme case, for $k \approx n$ we already have a constant fraction of the mitigated cases. This scenario appears in \cite{glos2020space}. So we can say it might be infeasible to eliminate an error for a single number, but, it still may be beneficial for multiple registers.

Note that this approach can also be used for one-hot encoding in combination with post-selection through compression scheme discussed in Subsection 3.2. One can check if the compressed number in binary is representing a number greater than or equal to $ n $ in an $ n $-qubit circuit. In this case, the depth and the number of gates will be $ O(n) $.

In addition, the proposed approach can be applied to Gray-code encoding \cite{sawaya2020resource}, after transforming it to binary encoding using the circuit given in Fig.~\ref{fig:wall-to-onehot}. The transformation has no impact on any of the resource measures.

\subsection{One-hot and binary mixed}

Finally, let us consider a combination of one-hot encoding and binary encoding proposed in~\cite{glos2020space,sawaya2020resource}. In such cases, bits encoding a single number are partitioned into $l$ groups, each group consisting of $ m $ qubits, and only one of the groups has nonzero bits. If $\bar l$-th group is the one with nonzero bits, then the bits of $\bar l$-th group encodes the number $x_{\bar l}$ in binary or Gray-code encoding, and the value of the encoded number is $(\bar l-1) (2^m-1)+ x_{\bar l}-1$. For example for $4$ groups, each with 2 bits, for the sequence $00\,00\,10\,00$ we have $\bar l=3$ and thus the value encoded is $(3-1) (2^2-1)+10_2-1=7$. Note that two conditions can be asserted: Exactly one group consists of nonzero bits,  and the last group may only attain some values due to the redundancy of binary encoding described in the previous paragraph. The latter can be solved the same way as it was solved for purely binary encoding in Sec. \ref{sec:binarytogrey}. For the former, we need to check whether the number of groups in which all consecutive bits in the group are all zeros is equal to $ l-1 $. 

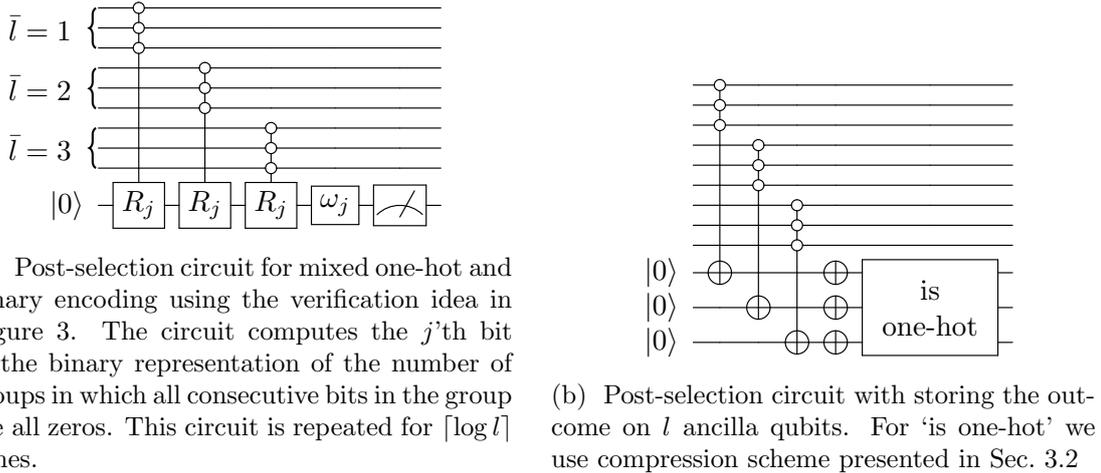
\begin{figure} \centering

	\begin{subfigure}[b]{0.45\textwidth}\centering
		\centering
		\begin{tikzpicture}
		\node at (0,0) {\mbox{\Qcircuit @C=0.5em @R=.3em {
					&&&&				& \ctrlo{1} 		& \qw 		& \qw 		& \qw				&\qw	&\qw  \\
					\bar{l}=1	&&&&				& \ctrlo{1} 		& \qw 		& \qw 		& \qw				&\qw	&\qw  \\
					&&&&				& \ctrlo{7} 		& \qw 		& \qw 		& \qw 				&\qw	&\qw \\
					&&&&				& \qw 				& \ctrlo{1}	& \qw 		& \qw 				&\qw	&\qw \\
					\bar{l}=2	&&&&				& \qw 				& \ctrlo{1} & \qw 		& \qw 				&\qw	&\qw \\
					&&&&				& \qw 				& \ctrlo{4} & \qw 		& \qw 				&\qw	&\qw \\
					&&&&				& \qw 				& \qw 		& \ctrlo{1} & \qw 				&\qw	&\qw \\
					\bar{l}=3	&&&&				& \qw 				& \qw 		& \ctrlo{1} & \qw 				&\qw	&\qw \\
					&&&&				& \qw 				& \qw 		& \ctrlo{1} & \qw 				&\qw	&\qw \\
					&&&&\lstick{\ket{0}}& \gate{R_j}		& \gate{R_j}& \gate{R_j}& \gate{\omega_j}	&\meter	&\qw \\
		}}};	
		\draw[decorate,decoration = {brace}, thick] (-1.9, 0.89) --  (-1.9, 1.42);
		\draw[decorate,decoration = {brace}, thick] (-1.9, 0.09) --  (-1.9, 0.62);
		\draw[decorate,decoration = {brace}, thick] (-1.9, -0.7) --  (-1.9, -0.17);
		\end{tikzpicture}
		\subcaption{\label{fig:binary-onehot-a} Post-selection circuit for mixed one-hot and binary encoding using the verification idea in Figure \ref{fig:sum-rotation}. The circuit computes the $ j $'th bit of the binary representation of the number of groups in which all consecutive bits in the group are all zeros. This circuit is repeated for $ \lceil \log l \rceil $ times. }
	\end{subfigure}
		\hspace{0.1in}
		\begin{subfigure}[b]{0.45\textwidth}\centering
			\centering
			\begin{tikzpicture}
			\node at (0,0) {\mbox{\Qcircuit @C=0.5em @R=.3em {
						&&&&				& \ctrlo{1} 		& \qw 		& \qw 		& \qw				&\qw	&\qw  \\
						&&&&				& \ctrlo{1} 		& \qw 		& \qw 		& \qw				&\qw	&\qw  \\
						&&&&				& \ctrlo{7} 		& \qw 		& \qw 		& \qw 				&\qw	&\qw \\
						&&&&				& \qw 				& \ctrlo{1}	& \qw 		& \qw 				&\qw	&\qw \\
						&&&&				& \qw 				& \ctrlo{1} & \qw 		& \qw 				&\qw	&\qw \\
						&&&&				& \qw 				& \ctrlo{5} & \qw 		& \qw 				&\qw	&\qw \\
						&&&&				& \qw 				& \qw 		& \ctrlo{1} & \qw 				&\qw	&\qw \\
						&&&&				& \qw 				& \qw 		& \ctrlo{1} & \qw 				&\qw	&\qw \\
						&&&&				& \qw 				& \qw 		& \ctrlo{3} & \qw 				&\qw	&\qw \\
						&&&&\lstick{\ket{0}}& \targ		& \qw& \qw& \targ & \multigate{2}{\parbox{1.4cm}{\centering is one-hot}}	&\qw \\
						&&&&\lstick{\ket{0}}& \qw		& \targ& \qw&\targ & \ghost{\parbox{1.4cm}{\centering is one-hot}}	&\qw \\
						&&&&\lstick{\ket{0}}& \qw		& \qw& \targ&\targ & \ghost{\parbox{1.4cm}{\centering is one-hot}}	&\qw \\
			}}};	
			\end{tikzpicture}
		\subcaption{\label{fig:binary-onehot-b} Post-selection circuit with storing the outcome on $l$ ancilla qubits. For `is one-hot' we use compression scheme presented in Sec.~\ref{sec:compression}  }
	\end{subfigure}
	\caption{\label{fig:extra-encoding} Post-selection circuits for binary encoding and mixed encoding.}
\end{figure}

One approach is to count the number of such groups using the verification idea from Fig.~\ref{fig:sum-rotation}. To implement the circuit, we need to implement a rotation gate controlled by $ m $ qubits for each one of the $ l $ groups. Using the ancilla-free and non-ancilla-free implementations of multi-controlled NOT gate, this would require $ O(lm^2) $ and $ O(lm) $ gates, respectively. Recall that there are two different approaches for verification, one using 1 ancilla qubit and the other using $ \log l $ qubits. Single-ancilla verification idea is visualized in Fig. \ref{fig:binary-onehot-a}. To save qubits, one may prefer ancilla-free multi-controlled NOT gate and single ancilla verification, overall which would require 1 ancilla qubit, $ O(lm^2\log l) $ gates and has $ O(lm\log l) $ depth. To have a circuit with smaller depth, one can use non-ancilla-free multi-controlled NOT gate and verification with $ \log l $ ancilla, resulting in $ O(m\log l) $ ancilla, $ O(lm \log l) $ gates and $ O(l \log m) $ depth.

In the second approach, the idea is to store the information whether each group consists of all zeros or not in an ancilla qubit. To implement this idea we use $l$ ancilla qubits, and save the required information. After applying NOTs on those qubits we can check whether the resulting $ l$-qubit state is an one-hot state. Then using the compression scheme for one-hot encoding from Sec. \ref{sec:compression}, we can check if the resulting state is one-hot. Using the ancilla-free implementation of multi-controlled NOT gate, this would require $ O(l) $ ancilla, $ O(lm^2) $ gates and $ O(l+m) $ depth. When we use non-ancilla-free implementation of multi-controlled NOT gate, then we have two options. We can use different ancilla for each multi-controlled NOT gate requiring $ O(lm) $ ancilla overall, and we can implement the circuit using $ O(l + \log m) $ depth, or using the same $ O(m) $ ancilla for each multi-controlled NOT gate, we can have a circuit with $ O(l+m) $ ancilla and $ O(l\log m ) $ depth. For both approaches, the number of required gates is $ O(lm) $.

This scheme can be also used for the mixture of Gray and one-hot encoding \cite{sawaya2020resource} after it is translated into binary encoding using the circuit given in Fig.~\ref{fig:wall-to-onehot}.

\subsection{Summary}

We present a summary of the resource requirements for the methods discussed so far. Depending on whether we use the 1- or $\log n$ ancilla counting method, we use notation $\Sigma_1$ and $\Sigma_{\log}$ respectively.  $ T_{\rm free} $ denotes the ancilla-free implementation of multi-controlled NOT gate, while $ T_{\rm ancilla} $ denotes the $O(n)$ ancilla implementation. $T_{\textrm{multi-anc}}$ is the case where it is guaranteed that for each implementation of the multi-controlled NOT gate, different ancilla qubits are available. Finally, $M$ denotes more subtle implementation details. For wall-domain encoding, it differentiates between the parallel checking with $O(n)$ and $1$ ancilla. For the mixed encoding, $M_{\rm store}$ denotes the approach of constructing one-hot vector and using the compression scheme. For all of the encodings, we assume that the analyzed system consists of $n$ qubits. For the mixed encoding, this also gives an identity $n=lm$.

In addition to the resources considered in the previous sections, we also present the volume of the encoding. The volume is defined as the product of depth and the number of qubits. For the number of qubits, we used the sum of ancilla qubits and $n$.
\renewcommand{\arraystretch}{1.2}
\begin{table}[h]
	\centering{ \small
\begin{tabular}{|l|c|ccc|c|}

	\hline
	\textbf{Encoding} &\textbf{Info.} & \textbf{Ancilla} & \textbf{Gates} & \textbf{Depth}  & \textbf{Volume}\\ \hline
	\multirow{2}*{\textbf{$k$-hot}} & $\Sigma_1$ \cite{mcardle2019error} & $1$ & $O(n \log n)$ & $O(n \log n)$ & $O(n^2 \log n)$  \\ \cline{2-6}
	& $\Sigma_{\log}$ & $O( \log n)$ & $O(n \log n)$ & $O(n)$ & $O(n^2)$ \\ \hline
	\multirow{2}*{\textbf{Domain-wall}} & $M_{\textrm{inductive}}$ & $1$ & $O(n )$ & $O(n)$ & $O(n^2)$ \\ \cline{2-6}
	& $M_{\rm parallel}$ & $O(n)$ & $O(n)$ & $O(1)$ & $O(n^3)$ \\ \hline
	\multirow{2}*{\textbf{Binary/Gray}} & $T_{\rm free}$ & 1 & $O(n^3)$ & $O(n^2)$  & $O(n^3)$ \\ \cline{2-6}
	& $T_{\rm anc}$ & $O(n)$ & $O(n^2)$ & $O(n\log n)$ & $O(n^2\log n)$\\\cline{2-6}
	\hline
	\multirow{6}*{\textbf{Mixed}} 	& $\Sigma_1 T_{\rm free} $ & $O(1)$ & $O(lm^2 \log l)$ & $O(lm \log l)$ & $O(l^2m^2 \log l)$ \\ \cline{2-6}
	& $\Sigma_1 T_{\rm anc}$ & $O(m)$ & $O( lm \log l)$ & $O( l \log l \log m)$ & $O(l^2 m\log l\log m )$ \\  \cline{2-6}
	& $\Sigma_{\log}T_{\rm free}$ & $ O(\log l) $ & $O( lm^2 \log l)$ & $O(lm)$ & $O(l^2m^2)$  \\ \cline{2-6}
	& $\Sigma_{\log}T_{\rm anc}$ & $O(m \log l) $ & $O( lm \log l)$ & $O(l \log m)$ & $O(l^2m\log m)$\\ \cline{2-6}
		& $M_{\rm store}T_{\rm free}$ & $ O(l) $ & $O( lm^2)$ & $O(l+m)$ & $O(l^2m+lm^2)$  \\ \cline{2-6}
	& $M_{\rm store}T_{\rm anc}$ & $O(l +  m) $ & $O( lm)$ & $O(l\log m)$ & $O(l^2m\log m) $\\
	\cline{2-6}
	& $M_{\rm store}T_{\textrm{multi-anc}}$ & $O(lm) $ & $O( lm)$ & $O(l^2m + \log m)$ & $O(l^2m +  lm\log m)$ \\ \hline
\end{tabular}}
\caption{Summary of the resource requirements of the post-selection circuits for different encoding using filtering.}
\end{table}
\begin{table}[h]
	\centering
\begin{tabular}{|l|cc|}
	\hline
	\textbf{Encoding} & \textbf{Gates} & \textbf{Depth}\\ \hline
	\textbf{$1$-hot} &  $O(n )$ & $O(n)$ \\ \hline
	\textbf{Domain-wall} & $O(n  )$ & $O(n )$  \\ \hline
\end{tabular}
\caption{Summary of the resource requirements of the post-selection circuits for different encodings using compression.}
\end{table}

\section{Application to Quantum Alternating Operator Ansatz}\label{sec:app}

Combinatorial optimization problems deal with minimizing or maximizing a function defined over a discrete set.
Quantum computing offers new approaches for solving such problems. As the first step, the problem should be expressed using a 2-local Ising model 
\begin{equation}
H = -\sum_{i>j}J_{ij}Z_iZ_j-\sum_jh_jZ_j,
\end{equation} 
whose ground state encodes the solution to the problem, where $Z_i$ is the Pauli-$ Z $ operator acting on the $i$-th qubit corresponding to spin variable $ s_i \in \{-1,1\} $, $ J_{ij} $ are the pairwise couplings, and $ h_j $ are the external magnetic fields. Then, the ground state can be approximated by quantum optimization algorithms like Quantum Annealing or Quantum Approximate Optimization Algorithm (QAOA). 

QAOA introduced by Farhi \textit{et.al} \cite{farhi2014quantum} finds an approximation to the ground state of $H$ by constructing a specific variational ansatz through first order Suzuki-Trotter decomposition approximating adiabatic evolution. The operators $\textrm{exp}\left(-\ii{\textit{r}}H_\textrm{mix}\right) $ and $ \textrm{exp}\left(-\ii{\textit{p}}H\right)$ are applied in alternation resulting in the state 
\begin{eqnarray}
\lvert\textit{\textbf{p}},\textbf{\textit{r}}\rangle =
\prod_{i = 1}^{l}\textrm{exp}\left(-\ii{\textit{r}}_iH_\textrm{mix}\right)\textrm{exp}\left(-\ii{\textit{p}}_iH\right)\lvert+\rangle^{\otimes n},
\end{eqnarray}
where  the initial state $\lvert+\rangle$ is the eigenstate of $X$, and $H_\textrm{mix} = -\sum_iX_i$. For a fixed number of layers $l$, QAOA requires $2l$ parameters i.e. $\textbf{\textit{r}} = \left(\textrm{\textit{r}}_1,\ldots ,  \textrm{\textit{r}}_l\right)$, $\textbf{\textit{p}} = \left(\textrm{\textit{p}}_1,\ldots , \textrm{\textit{p}}_l\right)$. The expectation value
$E_{\textbf{\textit{p}}, \textbf{\textit{r}} } = \langle \textbf{\textit{p}}, \textbf{\textit{r}} \lvert H\rvert \textbf{\textit{p}}, \textbf{\textit{r}}\rangle$,
of state $\lvert\textit{\textbf{p}},\textbf{\textit{r}}\rangle$ is approximated through measuring the state in the computational basis. The parameters $\textbf{\textit{p}}$, $\textbf{\textit{r}}$ are updated using classical procedures so that the energy $E_{\textbf{\textit{p}}, \textbf{\textit{r}} }$ is minimized. 

As long as both the objective and mixing Hamiltonian can be implemented efficiently, which is true for the 2-local Ising model and the given mixer Hamiltonian, QAOA can be used for any combinatorial problem. Many studies have been performed to characterize the properties of QAOA in the past few years. Rigorous proofs of computational power and reachability properties have been discussed \cite{morales2020universality,lloyd2018quantum,hastings2019classical,farhi2020quantum}, as well as characterization through heuristics, numerical experiments, and extensions of QAOA is introduced \cite{akshay2020reachability,zhu2020adaptive,wierichs2020avoiding}. QAOA has applications in a class of problems such as Max-Cut \cite{farhi2001quantum}, MaxE3Lin2 \cite{farhi2015quantum}, Max-$k$-Vertex Cover \cite{cook2020quantum}, sampling from Gibbs states \cite{verdon2017quantum}, and integer factorization \cite{anschuetz2019variational}. 

Quadratic Unconstrained Binary Optimization (QUBO) is a NP-Hard problem class, where optimization is done over binary variables $x_i\in \{0,1\}$, instead of spin variables $s_i \in\{-1,1\}$. QUBO is defined as
\begin{equation}
\sum_{i\leq j} x_i Q_{ij} x_j,
\end{equation}
where $ Q $ is a real matrix of coefficients defining the optimization problem. It is often more suitable to express a combinatorial optimization problem over binary variables using QUBO formulation, and the transformation between QUBO and Ising model can be performed easily using the mapping $x_i \leftrightarrow \frac{1-s_i}{2}$. 

In this paper, we will consider the Travelling Salesman Problem (TSP). QUBO formulation for TSP over $N$ cites is given as 
\begin{equation}
A\sum_{t=1}^{N}\left(1-\sum_{i=1}^{N}b_{t,i}\right)^2+A\sum_{i=1}^{N}\left(1-\sum_{t=1}^{N}b_{t,i}\right)^2+\sum_{\substack{i,j=1\\ i\ne j}}^{N}W_{ij}\sum_{t=1}^{N}b_{t,i}b_{t+1,j},
\label{eq:tsp-qubo}
\end{equation}
where $W$ is the cost matrix, and $b_{t,i}$, is the binary variable such that $b_{t,i}=1$ iff the $i$-th city is visited at time $t$ \cite{lucas2014ising}. $ A $  is a constant which needs to be adjusted so that the optimal solution of QUBO encodes the optimal solution for TSP. The formulation uses $ N^2 $ qubits which produces a large infeasible space $\ie$ there are $2^{N^2}$ possible solutions to QUBO model, while the number of routes is only $ N! = 2^{\order{N \log N}} $. To reduce the infeasible space, one possible approach is to encode the problem using less number of qubits as proposed in \cite{glos2020space}. Another approach is to reduce the effective space of the evolution, which is the idea behind the Quantum Alternating Operator Ansatz (QAOA+) \cite{hadfield2019quantum}.

QAOA+ is considered as an extension of QAOA that allows more general families of mixing operators. In QAOA+, the initial state is usually a feasible solution to the problem, and the mixer operator restricts the search to the feasible subspace by mapping feasible states to feasible states. Hence, the evolution takes place in a smaller subspace of the full Hilbert space, unlike QAOA. In this paper, we will consider a special case of QAOA+ called XY-QAOA. In XY-QAOA, the mixer is chosen as XY-Hamiltonian
\begin{equation}
\sum_{i=1}^NX_iX_{i+1}+Y_iY_{i+1},
\end{equation}
applied on every one-hot register, 
which preserves the Hamming weight of the quantum states \cite{wang2020x}. In the case of TSP over $ N $ cities, $ N $ registers each with $ N $ qubits are used such that if $ b_{t,i} $ = 1, then register $ t $ encodes $ i $ using one-hot encoding. The initial state can be prepared as the Kronecker product of $ W $-states which can be efficiently implemented \cite{wang2020x}. Note that the choice of the initial state is particularly suitable for XY-mixer as XY-mixer maps one-hot states to one-hot states. Although the generated subspace contains some infeasible states as well, it contains the whole feasible space for the TSP problem and is significantly smaller than the full Hilbert space. More precisely, the evolution takes place in $ N^N = 2^{\order{N\log N}} $ dimensional subspace of the full $ N^2 $-qubit Hilbert space.

As the post-selection scheme, we use the compression scheme for one-hot vectors proposed in Sec.~\ref{sec:compression}. We consider a noise model where every gate is affected by a random unitary channel applied after each quantum gate, including gates from the post-selection. In particular, we will consider depolarizing noise, amplitude damping noise, and random $X$ noise with parameter $\gamma$ reflecting the strength of the noise: the smaller the value of gamma, the least is the effect of the noise on the evolution. We assume that the initial state is $\ket{0\cdots 0}$ and measurements (both final and in the middle) are implemented perfectly. Ideal initial state preparation is justified as any digression into infeasible subspace will be detected by mid-circuit post-selection, or will produce some bias for QAOA+, which may be corrected by adjusting the parameters of the ansatz. For measurements, we note that the noise is highly biased. States $\ket{0}$ are much less prone to error compared to $\ket{1}$ so that it is unlikely that 1 is measured when one is expecting to measure 0~\cite{maciejewski2020mitigation}.

\begin{figure}
	\newcommand{\vlabeldagger}{\parbox{1.8cm}{\centering one-hot to binary}}
	\newcommand{\vlabel}{\parbox{1.8cm}{\centering binary to one-hot}}
	\newcommand{\biggatetext}{\parbox{1.8cm}{\centering QAOA\\layers}}
	\centering
	\mbox{	\Qcircuit @C=0.8em @R=0.1ex{%
			& \multigate{10}{\biggatetext} & \qw & \multigate{2}{\vlabeldagger} & \meter & \multigate{2}{\vlabel} & \qw \\
			& \ghost{\biggatetext} & \qw & \ghost{\vlabeldagger} & \qw & \ghost{\vlabel} & \qw \\
			& \ghost{\biggatetext} & \qw & \ghost{\vlabeldagger} & \qw & \ghost{\vlabel} & \qw \\
			& \nghost{\biggatetext} &  {} & {} & {} & {} & {} \\
			& \ghost{\biggatetext} & \qw & \multigate{2}{\vlabeldagger} & \meter & \multigate{2}{\vlabel} & \qw \\
			& \ghost{\biggatetext} & \qw & \ghost{\vlabeldagger}& \qw & \ghost{\vlabel} & \qw \\
			& \ghost{\biggatetext} & \qw & \ghost{\vlabeldagger}& \qw & \ghost{\vlabel} & \qw \\
			& \nghost{\biggatetext} & {} & {} & {} & {} & {} \\
			& \ghost{\biggatetext} & \qw & \multigate{2}{\vlabeldagger}& \meter & \multigate{2}{\vlabel} & \qw \\
			& \ghost{\biggatetext} & \qw & \ghost{\vlabeldagger} & \qw & \ghost{\vlabel} & \qw \\
			& \ghost{\biggatetext} & \qw & \ghost{\vlabeldagger}& \qw & \ghost{\vlabel} & \qw \\
	}}
	\caption{Illustration of the QAOA+ and mid-circuit post-selection scheme through compression. A brief discussion of the circuit components is given in Appendix \ref{sec:qaoa-details}.}
	\label{fig:simplified-qaoa}
\end{figure}
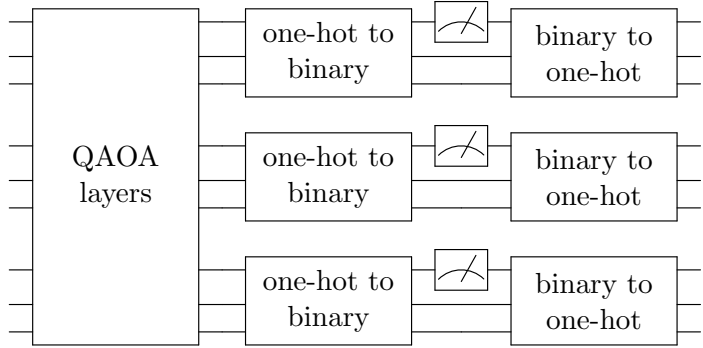

A simplified version of the circuit is visualized in Fig. [\ref{fig:simplified-qaoa}]. After a fixed number of QAOA layers we apply the compression scheme presented in Sec.~\ref{sec:compression}. We continue computation iff all measurements result in 0 states. For the final measurements, we post-select only those measurement samples which would appear in the error-robust computation. Note that in fact the mid-circuit post-selection can be also applied in the middle of the objective Hamiltonian application--this Hamiltonian is implemented by consecutively applying diagonal matrices, which doesn't change the space over which the states is defined. However, we apply mid-circuit post-selection after at the end of layers only for simplicity.

\begin{figure}[tbh!]
	\centering
	\includegraphics[]{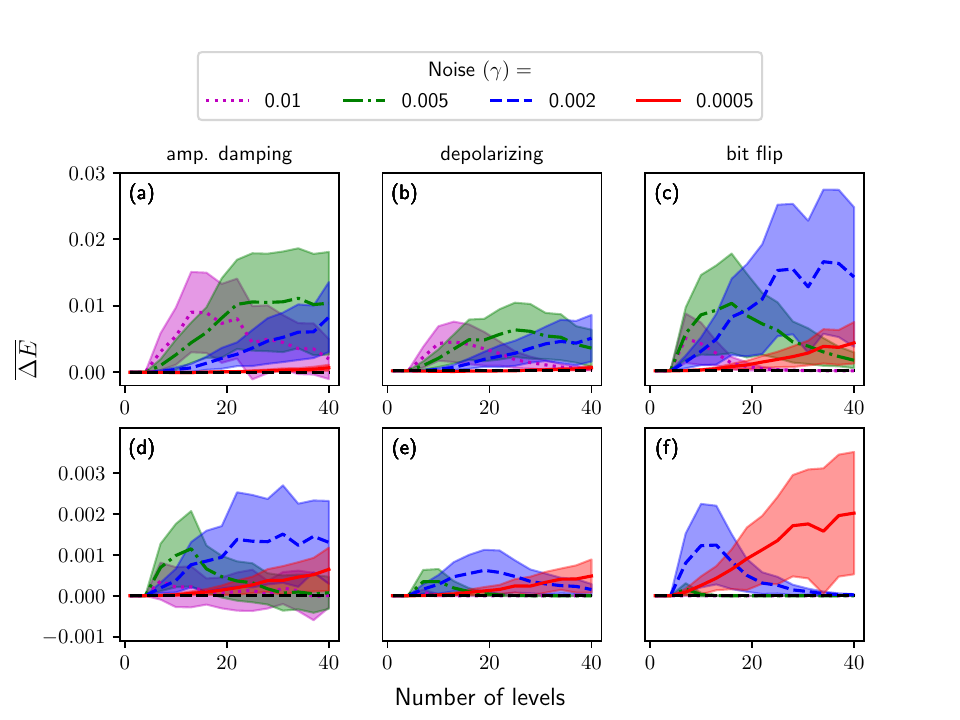}
	\caption{The efficiency of mid-circuit postselection against final circuit postselection only. The subplots (a), (b) and (c) are for $3$ cities, and (d), (e), (f) are showing the results for $4$ cities.}
	\label{fig:xy-rand}
\end{figure}

\begin{figure}[tbh!]
	\centering
	\includegraphics[]{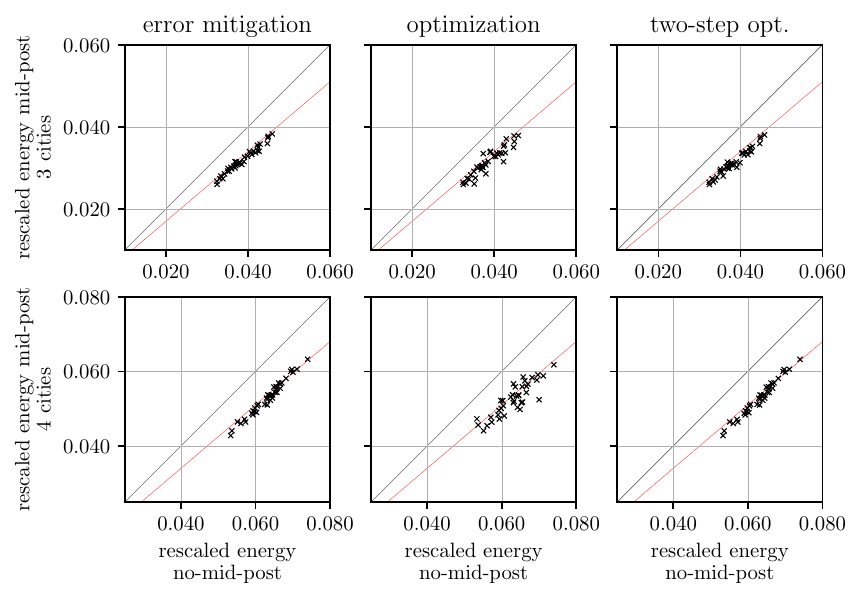}
		\caption{Effect of post-selection on QAOA optimization. In the first column, we simply correct the optimal angles obtained through regular optimization with the post-selection applied at every 2 layers. In the second column, we compare optimization with and without mid-circuit post-selection, starting with the same angles. Finally, in the last column, we take the optimal angles obtained through regular optimization and repeat the optimization with mid-circuit post-selection. The solid black line is the $y=x$ and denotes the `no difference' case. The red solid line is the $y=0.85x$.}
			\label{fig:optimization}
\end{figure}

We start by investigating the effect of the post-selection for randomly chosen angles. We sample 100 instances of TSP for $n=3,4$ cities. Cost matrix $W$ is a random matrix with elements sampled i.i.d. from the range $\{1,\dots,9\}$. The penalty value equals $A=2\max _{i,j}W_{ij}$. In the rest of the discussion, any considered energy will be for rescaled QUBO, such that the corresponding (attainable) maximal value of the pseudo-Boolean function is 1, and the smallest value is 0.  Let $E$ be the true energy coming from the noise-robust evolution and let $E_{\textrm{no-mid}}$ be the energy obtained from the noisy evolution but with post-selection applied on the final outcomes only. Finally, let $E_{\rm mid}$ be the energy with the post-selection applied both in the middle (at every 4th layer) and on the final outcomes. Our measure of quality is $\Delta E^{(i)} = |E^{(i)}-E_{\textrm{no-mid}}^{(i)}| - |E^{(i)}-E_{\rm mid}^{(i)}|$, where $i$ stands from the $i$-th TSP instance. Note that the larger the value, the more positive impact the mitigation scheme has on the output. 

The results presented in Fig.~\ref{fig:xy-rand} show that the effect of post-selection strongly depends not only on the noise impact $\gamma$, but also on the type of the noise. This is expected, as the noise structure also affects the way the amplitude is transferred between valid and invalid states. For example, for random $Z$ noise our method (the same as the classical postselection) cannot detect any deviation. However, for all combinations of noise strength and noise models, we see that the post-selection has mostly a positive effect on the evolution. 

The mean of $\Delta E$ is usually detached from zero. However, the area denoting the space within plus/minus standard deviation highly deviates from the mean. Yet, in most of the cases, the difference of the mean and standard deviation is close to 0, which shows that our method has likely no negative effect against final post-selection only. 

Let us now consider the effect of post-selection on the optimization process. We considered 40 TSP instances generated as described above, with 8 layers and random $X$ noise with $\gamma=0.002$. We consider 3 scenarios here. In all of them, we use the classical post-selection of the final outcomes, as classical post-selection can be implemented efficiently using classical computing. In the first scenario, we optimize the circuit without mid-circuit post-selection. Then we inject inside this circuit a mid-circuit post-selection procedure and compare the obtained energy with the previous one. This is the most efficient method, as the mid-circuit part of the circuit does not take part in the optimization process, which may slightly decrease the time required for the optimization. In the first column of Fig.~\ref{fig:optimization}, we can see that this approach provides stable improvement of around 15\% for both 3 and 4 cities case.

One may expect that correcting the energy via post-selection, through correcting the energy, will provide an alternative, more faithful energy landscape. To analyse this, we used mid-circuit post-selection also in the middle of optimization. We considered two approaches. In the first, we compare the energy obtained through circuits with and without post-selection, starting from the same initial angles. In the second, we first optimized the circuit without mid-circuit post-selection, and then we took the final circuit and re-optimized it with the circuit containing mid-circuit post-selection. Both are presented in Fig.~\ref{fig:optimization}. We can see that there is almost no difference between these two approaches, which indicates that the landscape may be very similar for these cases. This in turn implies, that it may be sufficient to optimize the circuit without the mid-circuit postselection, and only then apply the mid-circuit postselection.

\section{Conclusion}\label{sec:conc}

There have been some recent attempts to mitigate errors in Variational Quantum Algorithms (VQAs) through mid-circuit post-selection. Following this line of work, in this paper, we presented post-selection schemes for various encodings and different valid subspaces of quantum states, which can be used with VQA while solving particular combinatorial optimization problems and problems from quantum chemistry. We implemented the one-hot to binary post-selection through compression scheme to solve the Travelling Salesman Problem (TSP) using the Quantum Alternating Operator Ansatz (QAOA+) algorithm. The experiment results show that for amplitude damping, depolarizing, and bit-flip noises, the mid-circuit post-selection has a positive impact on the outcome compared to final post-selection only. The schemes we propose are qubit efficient, do not need classical operation, and use only mid-circuit measurements and reset. Hence, with the emerging technology of mid-circuit measurements [32, 33], the presented methods are currently applicable to NISQ algorithms.. Finally, our method can also be used in principle outside the scope of VQA. 

Although we have only considered the TSP problem in our numerical experiments, it is worth noting that the proposed schemes can be used with different objective Hamiltonians. Our ancilla-free post-selection through compression scheme can be applied to any problem where the feasible states are one-hot, including the problems defined over permutations such as Vehicle Routing Problem \cite{borowski2020new}, variations of TSP \cite{papalitsas2019qubo,salehi2021unconstrained}, Railway Dispatching Problem \cite{domino2021quadratic,domino2020quantum}, Graph Isomorphism Problem \cite{calude2017qubo}, Flight Gate Assignment Problem \cite{stollenwerk2019flight}.

There are several research directions that can be pursued requiring further investigation. First of all,  in general, the optimal number of post-selections to apply is not evident. Many factors should be considered here, including the complexity of the post-selection, the form of the feasible subspace $S$, the strength and form of the noise affecting the computation. It is desirable to design methods that would choose the optimal number (and perhaps the position) of mid-circuit post-selections to be applied. 

\paragraph{Acknowledgement}	
LB, AK, JAM, AG, \"OS have been partially supported by Polish National Science Center under the grant agreement 2019/33/B/ST6/02011. AG has been also supported by Polish National Science Center under the grant agreements 2020/37/N/ST6/02220. ZZ acknowledges support from the NKFIH Grants No.  K124152, FK135220, KH129601, K120569, and the Hungarian Quantum Technology National Excellence Program Project No. 2017-1.2.1-NKP-2017-00001 as well as from the Quantum Information National Laboratory of Hungary. We would like to thank Abuzer Yakaryılmaz for his valuable comments.

	\bibliographystyle{ieeetr}
	\bibliography{encoding_change}

\begin{thebibliography}{10}

\bibitem{preskill2018quantum}
J.~Preskill, ``{Quantum} computing in the {NISQ} era and beyond,'' {\em
  Quantum}, vol.~2, p.~79, 2018.

\bibitem{bharti2021noisy}
K.~Bharti, A.~Cervera-Lierta, T.~H. Kyaw, T.~Haug, S.~Alperin-Lea, A.~Anand,
  M.~Degroote, H.~Heimonen, J.~S. Kottmann, T.~Menke, {\em et~al.}, ``Noisy
  intermediate-scale quantum ({NISQ}) algorithms,'' {\em arXiv:2101.08448},
  2021.

\bibitem{kottmann2021tequila}
J.~Kottmann, S.~Alperin-Lea, T.~Tamayo-Mendoza, A.~Cervera-Lierta, C.~Lavigne,
  T.-C. Yen, V.~Verteletskyi, P.~Schleich, A.~Anand, M.~Degroote, {\em et~al.},
  ``{TEQUILA:} {A} platform for rapid development of quantum algorithms.,''
  {\em Quantum Science and Technology}, 2021.

\bibitem{Cerezo2021}
M.~Cerezo, A.~Arrasmith, R.~Babbush, S.~C. Benjamin, S.~Endo, K.~Fujii, J.~R.
  McClean, K.~Mitarai, X.~Yuan, L.~Cincio, and P.~J. Coles, ``{Variational
  quantum algorithms},'' {\em Nature Reviews Physics}, 2021.
\newblock DOI: 10.1038/s42254-021-00348-9.

\bibitem{peruzzo2014variational}
A.~Peruzzo, J.~McClean, P.~Shadbolt, M.-H. Yung, X.-Q. Zhou, P.~J. Love,
  A.~Aspuru-Guzik, and J.~L. O’{B}rien, ``A variational eigenvalue solver on
  a photonic quantum processor,'' {\em Nature Communications}, vol.~5, no.~1,
  pp.~1--7, 2014.

\bibitem{farhi2014quantum}
E.~Farhi, J.~Goldstone, and S.~Gutmann, ``A quantum approximate optimization
  algorithm,'' Tech. Rep. MIT-CTP/4610, 2014.

\bibitem{arute2020hartree}
F.~Arute, K.~Arya, R.~Babbush, D.~Bacon, J.~C. Bardin, R.~Barends, S.~Boixo,
  M.~Broughton, B.~B. Buckley, D.~A. Buell, {\em et~al.}, ``Hartree-{F}ock on a
  superconducting qubit quantum computer,'' {\em Science}, vol.~369, no.~6507,
  pp.~1084--1089, 2020.

\bibitem{mitarai2018quantum}
K.~Mitarai, M.~Negoro, M.~Kitagawa, and K.~Fujii, ``Quantum circuit learning,''
  {\em Physical Review A}, vol.~98, no.~3, p.~032309, 2018.

\bibitem{khatri2019quantum}
S.~Khatri, R.~LaRose, A.~Poremba, L.~Cincio, A.~T. Sornborger, and P.~J. Coles,
  ``Quantum-assisted quantum compiling,'' {\em Quantum}, vol.~3, p.~140, 2019.

\bibitem{moll2018quantum}
N.~Moll, P.~Barkoutsos, L.~S. Bishop, J.~M. Chow, A.~Cross, D.~J. Egger,
  S.~Filipp, A.~Fuhrer, J.~M. Gambetta, M.~Ganzhorn, {\em et~al.}, ``Quantum
  optimization using variational algorithms on near-term quantum devices,''
  {\em Quantum Science and Technology}, vol.~3, no.~3, p.~030503, 2018.

\bibitem{o2016scalable}
P.~J. O’Malley, R.~Babbush, I.~D. Kivlichan, J.~Romero, J.~R. McClean,
  R.~Barends, J.~Kelly, P.~Roushan, A.~Tranter, N.~Ding, {\em et~al.},
  ``Scalable quantum simulation of molecular energies,'' {\em Physical Review
  X}, vol.~6, no.~3, p.~031007, 2016.

\bibitem{mcclean2016theory}
J.~R. McClean, J.~Romero, R.~Babbush, and A.~Aspuru-Guzik, ``The theory of
  variational hybrid quantum-classical algorithms,'' {\em New Journal of
  Physics}, vol.~18, no.~2, p.~023023, 2016.

\bibitem{endo2021hybrid}
S.~Endo, Z.~Cai, S.~C. Benjamin, and X.~Yuan, ``Hybrid quantum-classical
  algorithms and quantum error mitigation,'' {\em Journal of the Physical
  Society of Japan}, vol.~90, no.~3, p.~032001, 2021.

\bibitem{bravyi2021mitigating}
S.~Bravyi, S.~Sheldon, A.~Kandala, D.~C. Mckay, and J.~M. Gambetta,
  ``Mitigating measurement errors in multiqubit experiments,'' {\em Physical
  Review A}, vol.~103, no.~4, p.~042605, 2021.

\bibitem{geller2020efficient}
M.~R. Geller and M.~Sun, ``Efficient correction of multiqubit measurement
  errors,'' {\em arXiv:2001.09980}, 2020.

\bibitem{maciejewski2021modeling}
F.~B. Maciejewski, F.~Baccari, Z.~Zimbor{\'{a}}s, and M.~Oszmaniec, ``Modeling
  and mitigation of cross-talk effects in readout noise with applications to
  the {Q}uantum {A}pproximate {O}ptimization {A}lgorithm,'' {\em {Quantum}},
  vol.~5, p.~464, 2021.

\bibitem{hadfield2019quantum}
S.~Hadfield, Z.~Wang, B.~O’Gorman, E.~G. Rieffel, D.~Venturelli, and
  R.~Biswas, ``From the {Q}uantum {A}pproximate {O}ptimization {A}lgorithm to a
  {Q}uantum {A}lternating {O}perator {A}nsatz,'' {\em Algorithms}, vol.~12,
  no.~2, p.~34, 2019.

\bibitem{wang2020x}
Z.~Wang, N.~C. Rubin, J.~M. Dominy, and E.~G. Rieffel, ``{$XY$} mixers:
  Analytical and numerical results for the quantum alternating operator
  ansatz,'' {\em Physical Review A}, vol.~101, no.~1, p.~012320, 2020.

\bibitem{bartschi2020grover}
A.~B{\"a}rtschi and S.~Eidenbenz, ``Grover mixers for {QAOA}: {S}hifting
  complexity from mixer design to state preparation,'' in {\em 2020 IEEE
  International Conference on Quantum Computing and Engineering (QCE)},
  pp.~72--82, IEEE, 2020.

\bibitem{gard2020efficient}
B.~T. Gard, L.~Zhu, G.~S. Barron, N.~J. Mayhall, S.~E. Economou, and E.~Barnes,
  ``Efficient symmetry-preserving state preparation circuits for the
  variational quantum eigensolver algorithm,'' {\em npj Quantum Information},
  vol.~6, no.~1, pp.~1--9, 2020.

\bibitem{ryabinkin2018constrained}
I.~G. Ryabinkin, S.~N. Genin, and A.~F. Izmaylov, ``Constrained variational
  quantum eigensolver: Quantum computer search engine in the {F}ock space,''
  {\em Journal of chemical theory and computation}, vol.~15, no.~1,
  pp.~249--255, 2018.

\bibitem{kerenidis2021classical}
I.~Kerenidis, J.~Landman, and N.~Mathur, ``Classical and quantum algorithms for
  orthogonal neural networks,'' {\em arXiv:2106.07198}, 2021.

\bibitem{mcardle2019error}
S.~McArdle, X.~Yuan, and S.~Benjamin, ``Error-mitigated digital quantum
  simulation,'' {\em Physical Review Letters}, vol.~122, no.~18, p.~180501,
  2019.

\bibitem{shaydulin2021error}
R.~Shaydulin and A.~Galda, ``Error mitigation for deep quantum optimization
  circuits by leveraging problem symmetries,'' {\em arXiv:2106.04410}, 2021.

\bibitem{streif2021quantum}
M.~Streif, M.~Leib, F.~Wudarski, E.~Rieffel, and Z.~Wang, ``Quantum algorithms
  with local particle-number conservation: Noise effects and error
  correction,'' {\em Physical Review A}, vol.~103, no.~4, p.~042412, 2021.

\bibitem{lucas2014ising}
A.~Lucas, ``Ising formulations of many {NP} problems,'' {\em Frontiers in
  Physics}, vol.~2, p.~5, 2014.

\bibitem{glos2020space}
A.~Glos, A.~Krawiec, and Z.~Zimbor{\'a}s, ``Space-efficient binary optimization
  for variational computing,'' {\em arXiv:2009.07309}, 2020.

\bibitem{tabi2020quantum}
Z.~Tabi, K.~H. El-Safty, Z.~Kallus, P.~H{\'a}ga, T.~Kozsik, A.~Glos, and
  Z.~Zimbor{\'a}s, ``Quantum optimization for the graph coloring problem with
  space-efficient embedding,'' in {\em 2020 IEEE International Conference on
  Quantum Computing and Engineering (QCE)}, pp.~56--62, IEEE, 2020.

\bibitem{tamura2021performance}
K.~Tamura, T.~Shirai, H.~Katsura, S.~Tanaka, and N.~Togawa, ``Performance
  comparison of typical binary-integer encodings in an {I}sing machine,'' {\em
  IEEE Access}, vol.~9, pp.~81032--81039, 2021.

\bibitem{fuchs2021efficient}
F.~G. Fuchs, H.~{\O}. Kolden, N.~H. Aase, and G.~Sartor, ``Efficient encoding
  of the weighted {MAX}-$k$-{CUT} on a quantum computer using {QAOA},'' {\em SN
  Computer Science}, vol.~2, no.~2, pp.~1--14, 2021.

\bibitem{chancellor2019domain}
N.~Chancellor, ``Domain wall encoding of discrete variables for quantum
  annealing and {QAOA},'' {\em Quantum Science and Technology}, vol.~4, no.~4,
  p.~045004, 2019.

\bibitem{IBM}
N.~Paul and B.~Johnson, ``How to measure and reset a qubit in the middle of a
  circuit execution,'' 2021.
\newblock
  \url{https://www.ibm.com/blogs/research/2021/02/quantum-mid-circuit-measurement/}.

\bibitem{honeywell}
``Honeywell {S}ystem model {H}1,'' 2020.
\newblock
  \url{https://www.honeywell.com/us/en/company/quantum/quantum-computer}.

\bibitem{georgescu2014quantum}
I.~M. Georgescu, S.~Ashhab, and F.~Nori, ``Quantum simulation,'' {\em Reviews
  of Modern Physics}, vol.~86, no.~1, p.~153, 2014.

\bibitem{bartschi2019deterministic}
A.~B{\"a}rtschi and S.~Eidenbenz, ``Deterministic preparation of {D}icke
  states,'' in {\em International Symposium on Fundamentals of Computation
  Theory}, pp.~126--139, Springer, 2019.

\bibitem{cook2020quantum}
J.~Cook, S.~Eidenbenz, and A.~B{\"a}rtschi, ``The {Q}uantum {A}lternating
  {O}perator {A}nsatz on {M}aximum $k$-{V}ertex {C}over,'' in {\em 2020 IEEE
  International Conference on Quantum Computing and Engineering (QCE)},
  pp.~83--92, IEEE, 2020.

\bibitem{sawaya2020resource}
N.~P. Sawaya, T.~Menke, T.~H. Kyaw, S.~Johri, A.~Aspuru-Guzik, and G.~G.
  Guerreschi, ``Resource-efficient digital quantum simulation of $d$-level
  systems for photonic, vibrational, and spin-$s$ hamiltonians,'' {\em npj
  Quantum Information}, vol.~6, no.~1, pp.~1--13, 2020.

\bibitem{karimi2019practical}
S.~Karimi and P.~Ronagh, ``Practical integer-to-binary mapping for quantum
  annealers,'' {\em Quantum Information Processing}, vol.~18, no.~4, pp.~1--24,
  2019.

\bibitem{saeedi2013linear}
M.~Saeedi and M.~Pedram, ``Linear-depth quantum circuits for $n$-qubit
  {T}offoli gates with no ancilla,'' {\em Physical Review A}, vol.~87, no.~6,
  p.~062318, 2013.

\bibitem{he2017decompositions}
Y.~He, M.-X. Luo, E.~Zhang, H.-K. Wang, and X.-F. Wang, ``Decompositions of
  $n$-qubit {T}offoli gates with linear circuit complexity,'' {\em
  International Journal of Theoretical Physics}, vol.~56, no.~7,
  pp.~2350--2361, 2017.

\bibitem{morales2020universality}
M.~E. Morales, J.~D. Biamonte, and Z.~Zimbor{\'a}s, ``On the universality of
  the quantum approximate optimization algorithm,'' {\em Quantum Information
  Processing}, vol.~19, no.~9, pp.~1--26, 2020.

\bibitem{lloyd2018quantum}
S.~Lloyd, ``Quantum approximate optimization is computationally universal,''
  {\em arXiv:1812.11075}, 2018.

\bibitem{hastings2019classical}
M.~B. Hastings, ``Classical and quantum bounded depth approximation
  algorithms,'' {\em arXiv:1905.07047}, 2019.

\bibitem{farhi2020quantum}
E.~Farhi, D.~Gamarnik, and S.~Gutmann, ``The quantum approximate optimization
  algorithm needs to see the whole graph: Worst case examples,'' {\em
  arXiv:2005.08747}, 2020.
\newblock Technical report MIT-CTP/5206.

\bibitem{akshay2020reachability}
V.~Akshay, H.~Philathong, M.~E. Morales, and J.~D. Biamonte, ``Reachability
  deficits in quantum approximate optimization,'' {\em Physical review
  letters}, vol.~124, no.~9, p.~090504, 2020.

\bibitem{zhu2020adaptive}
L.~Zhu, H.~L. Tang, G.~S. Barron, F.~Calderon-Vargas, N.~J. Mayhall, E.~Barnes,
  and S.~E. Economou, ``An adaptive quantum approximate optimization algorithm
  for solving combinatorial problems on a quantum computer,'' {\em
  arXiv:2005.10258}, 2020.

\bibitem{wierichs2020avoiding}
D.~Wierichs, C.~Gogolin, and M.~Kastoryano, ``Avoiding local minima in
  variational quantum eigensolvers with the natural gradient optimizer,'' {\em
  Physical Review Research}, vol.~2, no.~4, p.~043246, 2020.

\bibitem{farhi2001quantum}
E.~Farhi, J.~Goldstone, S.~Gutmann, J.~Lapan, A.~Lundgren, and D.~Preda, ``A
  quantum adiabatic evolution algorithm applied to random instances of an
  {NP}-complete problem,'' {\em Science}, vol.~292, no.~5516, pp.~472--475,
  2001.

\bibitem{farhi2015quantum}
E.~Farhi, J.~Goldstone, and S.~Gutmann, ``A quantum approximate optimization
  algorithm applied to a bounded occurrence constraint problem,'' {\em
  arXiv:1412.6062}, 2014.
\newblock Technical Report MIT-CTP/4628.

\bibitem{verdon2017quantum}
G.~Verdon, M.~Broughton, and J.~Biamonte, ``A quantum algorithm to train neural
  networks using low-depth circuits,'' {\em arXiv:1712.05304}, 2017.

\bibitem{anschuetz2019variational}
E.~Anschuetz, J.~Olson, A.~Aspuru-Guzik, and Y.~Cao, ``Variational quantum
  factoring,'' in {\em International Workshop on Quantum Technology and
  Optimization Problems}, pp.~74--85, Springer, 2019.

\bibitem{maciejewski2020mitigation}
F.~B. Maciejewski, Z.~Zimbor{\'a}s, and M.~Oszmaniec, ``Mitigation of readout
  noise in near-term quantum devices by classical post-processing based on
  detector tomography,'' {\em Quantum}, vol.~4, p.~257, 2020.

\bibitem{borowski2020new}
M.~Borowski, P.~Gora, K.~Karnas, M.~B{\l}ajda, K.~Kr{\'o}l, A.~Matyjasek,
  D.~Burczyk, M.~Szewczyk, and M.~Kutwin, ``New hybrid quantum annealing
  algorithms for solving {V}ehicle {R}outing {P}roblem,'' in {\em International
  Conference on Computational Science}, pp.~546--561, Springer, 2020.

\bibitem{papalitsas2019qubo}
C.~Papalitsas, T.~Andronikos, K.~Giannakis, G.~Theocharopoulou, and
  S.~Fanarioti, ``A {QUBO} model for the {T}raveling {S}alesman {P}roblem with
  {T}ime {W}indows,'' {\em Algorithms}, vol.~12, no.~11, p.~224, 2019.

\bibitem{salehi2021unconstrained}
{\"O}.~Salehi, A.~Glos, and J.~A. Miszczak, ``Unconstrained binary models of
  the {T}ravelling {S}alesman {P}roblem variants for quantum optimization,''
  {\em arXiv:2106.09056}, 2021.

\bibitem{domino2021quadratic}
K.~Domino, A.~Kundu, and K.~Krawiec, ``Quadratic and higher-order unconstrained
  binary optimization of railway dispatching problem for quantum computing,''
  {\em arXiv:2107.03234}, 2021.

\bibitem{domino2020quantum}
K.~Domino, M.~Koniorczyk, K.~Krawiec, K.~Ja{\l}owiecki, and B.~Gardas,
  ``Quantum computing approach to railway dispatching and conflict management
  optimization on single-track railway lines,'' {\em arXiv:2010.08227}, 2020.

\bibitem{calude2017qubo}
C.~S. Calude, M.~J. Dinneen, and R.~Hua, ``{QUBO} formulations for the graph
  isomorphism problem and related problems,'' {\em Theoretical Computer
  Science}, vol.~701, pp.~54--69, 2017.

\bibitem{stollenwerk2019flight}
T.~Stollenwerk, E.~Lobe, and M.~Jung, ``Flight gate assignment with a quantum
  annealer,'' in {\em International Workshop on Quantum Technology and
  Optimization Problems}, pp.~99--110, Springer, 2019.

\bibitem{cabello2002}
A.~Cabello, ``Bell's theorem with and without inequalities for the three-qubit
  {G}reenberger-{H}orne-{Z}eilinger and {W} states,'' {\em Phys. Rev. A},
  vol.~65, p.~032108, Feb 2002.

\end{thebibliography}

\appendix

\section{Details on numerical experiments}

For the simulations, we used \texttt{qiskit} programming framework. The code is available on \url{https://github.com/iitis/ec-qaoa-code}. Versions of the software used are available in the \texttt{.yml} file in the link.

\subsection{Noise model}

We considered a noise model which applies a noise channel after each gate on the qubits on which the gate is acting on. For a noise parameter $\gamma$ and for different noise models, the channels are expressed as follows.
\begin{enumerate}
	\item Depolarizing channel for 1- and 2-qubit gates:
	\begin{equation}
	\mathcal N_D(\varrho) \coloneqq (1-\gamma)\varrho + \gamma \frac{\Id}{\dim(\varrho )}.
	\end{equation}
	\item Amplitude damping channel for 1-qubit gates:
	\begin{equation}
	\mathcal N_A(\varrho) \coloneqq \begin{bmatrix}
	1 & 0 \\ 0 & \sqrt {1-\gamma}
	\end{bmatrix} \varrho \begin{bmatrix}
	1 & 0 \\ 0 & \sqrt {1-\gamma}
	\end{bmatrix} + \begin{bmatrix}
	0 & 0 \\ \sqrt \gamma & 0
	\end{bmatrix} \varrho \begin{bmatrix}
	0 & \sqrt{\gamma} \\ 0 & 0
	\end{bmatrix}.
	\end{equation}
	For 2-qubit gates, we took $\mathcal N_A^{\otimes 2}$.
	\item Random $X$ damping channel for 1-qubit gates
	\begin{equation}
	\mathcal N_X(\varrho) \coloneqq (1-\gamma)\varrho + \gamma X\varrho X.
	\end{equation}
	For 2-qubit gates we took $\mathcal N_X^{\otimes 2}$.
\end{enumerate}

To estimate the energy, we used the \texttt{density\_matrix} simulator, \ie so the energy estimation is always exact given the noise model. That is equivalent to estimating the energy with infinitely many samples with the same machine. 

All circuits were transpiled so that they only consist of general 1-qubit gates and controlled-NOTs. We assumed full connectivity for the simulator, meaning that the controlled-NOT can be implemented between any qubits in both directions.

\subsection{TSP}
We consider TSP from \cite{lucas2014ising} in a form
\begin{equation}
A\sum_{t=1}^{N}\left(1-\sum_{i=1}^{N}b_{t,i}\right)^2+A\sum_{i=1}^{N}\left(1-\sum_{t=1}^{N}b_{t,i}\right)^2+B\sum_{\substack{i,j=1\\ i\ne j}}^{N}W_{ij}\sum_{t=1}^{N}b_{t,i}b_{t+1,j},
\end{equation}
with $A = 2\max_{i,j}W_{i,j}$ and $B=1$. We considered $W$ to be a random matrix with entries chosen i.i.d. uniformly from $\{1,\dots,10\}$. We simplified the sampled TSP w.l.o.g. by assuming that the first city is visited in time 1, which dropped the qubits requirements from $N^2$ to $(N-1)^2$. Note the layout will be the same as for TSP with $N-1$ cities, so the QAOA+ algorithms used in the paper can be used.

\subsection{XY-QAOA} \label{sec:qaoa-details}
%

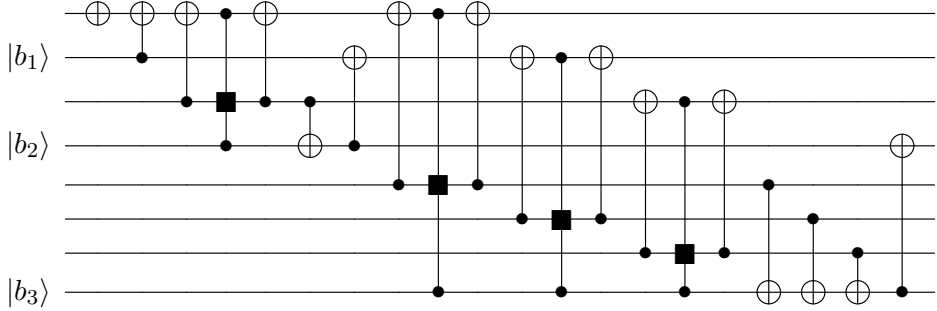
\begin{figure}
	\centering
		\mbox{\Qcircuit @C=0.7em @R=0.7em {
			& \targ & \targ     & \targ     & \ctrl{2}  & \targ 	& \qw		& \qw		&\targ 		&\ctrl{4}	&\targ		&\qw		&\qw		&\qw		&\qw		&\qw		&\qw		&\qw		&\qw	&\qw	&\qw&\qw\\
\lstick{\ket{b_1}}	& \qw   & \ctrl{-1} & \qw 		& \qw		& \qw		& \qw		& \targ		&\qw		& \qw		& \qw		&\targ		&\ctrl{4}	&\targ		&\qw		&\qw		&\qw		&\qw		&\qw	&\qw	&\qw&\qw\\
			& \qw   & \qw 	    & \ctrl{-2} & \targrel	& \ctrl{-2}	& \ctrl{1}	& \qw		&\qw		& \qw		& \qw		&\qw		&\qw		&\qw		&\targ		&\ctrl{4}	&\targ		&\qw		&\qw	&\qw	&\qw&\qw\\
\lstick{\ket{b_2}}	& \qw   & \qw		& \qw		& \ctrl{-1}	& \qw		& \targ		&\ctrl{-2}	&\qw		& \qw		& \qw		&\qw		&\qw		&\qw		&\qw		&\qw		&\qw		&\qw		&\qw	&\qw	&\targ&\qw\\
			& \qw   & \qw		& \qw		& \qw		& \qw		& \qw		&\qw		&\ctrl{-4}	&\targrel	& \ctrl{-4}	&\qw		&\qw		&\qw		&\qw		&\qw		&\qw		&\ctrl{3}	&\qw	&\qw	&\qw&\qw\\
			& \qw   & \qw		& \qw		& \qw		& \qw		& \qw		&\qw		&\qw		& \qw		& \qw		&\ctrl{-4}	&\targrel	&\ctrl{-4}	&\qw		&\qw		&\qw		&\qw		&\ctrl{2}&\qw	&\qw&\qw\\
			& \qw   & \qw		& \qw		& \qw		& \qw		& \qw		&\qw		&\qw		& \qw		& \qw		&\qw		&\qw		&\qw		&\ctrl{-4}	&\targrel	&\ctrl{-4}	&\qw		&\qw	&\ctrl{1}&\qw&\qw\\
\lstick{\ket{b_3}}	& \qw   & \qw		& \qw		& \qw		& \qw		& \qw		&\qw		&\qw		&\ctrl{-3}	& \qw	
			&\qw	&\ctrl{-2}	&\qw		&\qw		&\ctrl{-1}	&\qw		&\targ		&\targ	&\targ	&\ctrl{-4}&\qw\\
	}}
	\caption{Circuit for converting one-hot to binary encodings \cite{sawaya2020resource}. Multi-controlled black square stands for the relative phase Toffoli}
	\label{fig:ub-k-bu}
\end{figure}

For XY-QAOA, the mixer does not preserves the space of permutation states, but only of $N$ products of $N$-qubit one-hot vectors. Thus the algorithm starts in 
\begin{equation}
|W_N\rangle^{\otimes N}= \left(\frac{1}{\sqrt{N}}(|100 \ldots 0\rangle+|010 \ldots 0\rangle+\ldots+|00 \ldots 01\rangle)\right)^{\otimes N}.
\end{equation} 
The implementation for $\ket{W_N}$ can be found in \cite{wang2020x}.

A single layer of XY-QAOA is composed of a unitary $U_W$, which gives the initial state preparation, followed by the objective Hamiltonian, the mixer, and the unitary gate $V$ performing the post-selection measurement.
The initial state is given by the $\ket{W_N}$ state \cite{cabello2002} on each register indicating different time points.

The XY-QAOA mixer is a Trotter-Suzuki approximation of Hamiltonian $\sum_{i=1}^n(XY)_{i,i+1}$ with periodic condition $n+1\equiv1$ are implemented in order to minimize the circuit depth on each quantum register reflecting a single timepoint for TSP, where $(XY)_{i,j} = X_iX_j+Y_iY_j$. First, we implement all the gates for $XY_{i,i+1}$ with even $i$, following odd $i$ and for the last, considering periodic boundary conditions, $i=n$ and $i=1$. 

When implementing the objective Hamiltonian, we only included the part for computing the cost routes and for verifying whether at different time-points we have distinct cities. Note that the part which checks whether at given time point only one city is visited is guaranteed by the algorithm itself.

\subsection{Experiments}

\subsubsection{Energy difference}
We consider 100 TSP instances with penalty parameter $A=2\max_{i,j}W_{ij}$, where $W$ is the cost matrix with all elements sampled i.i.d. uniformly from the set $\{1,\dots,9\}$. For each TSP and layer we sampled a single angle vector with all elements sampled i.i.d. according to a uniform distribution over $[0,2\pi]$. We considered $1,\dots,40 $ number of layers. All energies were renormalized to the maximal $E_{\max}$ and minimal $E_{\min}$ energy of the Hamiltonian, i.e., $E\mapsto \frac{E-E_{\min}}{E_{\max} - E_{\min}}$. The mid-circuit postselection was done through the compression scheme, every 4th layer, while the final outcomes were filtered through the classical postselection. Algebraically it is equivalent to projection $\varrho \mapsto \Pi_S \varrho\Pi_S$. Note that mid-circuit postselection approach was compared to the case with the final postselection.

\subsubsection{Optimization}

For the optimization, we consider 8 layer QAOA for 40 TSP instances sampled as above (each considered once). We considered random $X$ noise with $\gamma=0.002$. Mid-circuit postselection was applied every second layer. For optimization, we chose the L-BFGS-B algorithm with default parameters from \texttt{scipy.optimize}. Initial angles were sampled as in the previous experiments.
\end{document}